\documentclass[10pt, twocolumn]{IEEEtran}
\usepackage{array,xcolor}
\usepackage{booktabs}
\usepackage{multirow}
\usepackage{comment}
\usepackage{setspace}

\renewenvironment{thebibliography}[1]{
  \begin{oldthebibliography}{#1}
    \setlength{\itemsep}{0em}
    \setlength{\parskip}{0em}
}
{
  \end{oldthebibliography}
}
\usepackage{lipsum}
\usepackage{bm}
\usepackage{tikz}
\usepackage{amsmath}
\usetikzlibrary{positioning, arrows.meta, shapes.geometric, decorations.pathreplacing, calc, fit, backgrounds}
\usepackage{xcolor}
\usetikzlibrary{positioning,calc,arrows.meta,fit,backgrounds}
\usepackage{pgfplots}
\usepackage{pgfplotstable}
\usetikzlibrary{spy}
    
\usepgfplotslibrary{colormaps}
\usepackage{mathrsfs}
\usepackage{mathtools}
\usetikzlibrary{shapes.geometric, arrows}
\usepgfplotslibrary{external}
\usepackage{slashbox}

\usepackage{pifont}
\pgfplotsset{grid style={dashed,gray}}
\pgfplotsset{minor grid style={dotted,gray}}
\pgfplotsset{major grid style={dashed,gray}}

\usepackage{caption}
\usepackage{subcaption}
\usepackage{amsmath, tabularx} 
\usepackage{hyperref}
\usepackage{cleveref}
\usepackage{algpseudocode}
\usepackage{ifpdf}
\usepackage{graphicx,amssymb,lineno}
\makeatletter
\newcommand*\bigcdot{\mathpalette\bigcdot@{.5}}
\newcommand*\bigcdot@[2]{\mathbin{\vcenter{\hbox{\scalebox{#2}{$\m@th#1\bullet$}}}}}
\makeatother
\usepackage{relsize}
\usepackage{filecontents}
\usepackage{lipsum}
\usepackage{color,soul}
\usepackage{algorithm}
\usepackage{color,soul}
\usepackage{cite}
\usepgfplotslibrary{groupplots}

\makeatletter
\def\BState{\State\hskip-\ALG@thistlm}
\makeatother
\makeatletter
\newcommand*{\rom}[1]{\expandafter\@slowromancap\romannumeral #1@}
\makeatother
\allowdisplaybreaks

\graphicspath{ {Images/} }

\usepackage[T1]{fontenc}
\usepackage{float}

\makeatletter
\newcommand{\multiline}[1]{%
  \begin{tabularx}{\dimexpr\linewidth-\ALG@thistlm}[t]{@{}X@{}}
    #1
  \end{tabularx}
}
\makeatother

\renewcommand{\arraystretch}{1.5}
\usepackage{changepage}
\usepackage{adjustbox}
\usepackage[printonlyused]{acronym}
\newacro{6g} [6G] {sixth generation}
\newacro{cr} [CR] {cognitive radio}
\newacro{isac} [ISAC] {integrated sensing and communication}
\newacro{sim} [SIM] {stacked intelligent metasurface}

\newacro{afdm} [AFDM] {affine frequency-division multiplexing}
\newacro{otfs} [OTFS] {orthogonal time–frequency space}
\newacro{cp} [CP] {cyclic prefix}
\newacro{idaft} [IDAFT] {inverse discrete affine fourier transform}
\newacro{daft} [DAFT] {discrete affine fourier transform}

\newacro{fft} [FFT] {fast Fourier transform}
\newacro{ifft} [IFFT] {inverse fast Fourier transform}

\newacro{sb} [SB] {secondary base station}
\newacro{su} [SU] {secondary user equipment}
\newacro{sws} [SWS] {secondary wireless system}
\newacro{pb} [PB] {primary base station}
\newacro{pu} [PU] {primary user equipment}
\newacro{pws} [PWS] {primary wireless system}
\newacro{ula} [ULA] {uniform linear array}
\newacro{csi} [CSI] {channel state information}
\newacro{miso} [MISO] {multiple input single output}
\newacro{ra} [RA] {reconfigurable antenna}
\newacro{los} [LoS] {line-of-sight}
\newacro{nlos} [NLoS] {non-line-of-sight}
\newacro{shod} [SHOD] {spherical harmonious orthogonal decomposition}
\newacro{ofdm} [OFDM] {orthogonal frequency division multiplexing}
\newacro{dof} [DoF] {degree of freedom}
\newacro{fim} [FIM] {Fisher information matrix}
\newacro{bfim} [BFIM] {Bayesian Fisher information matrix}
\newacro{em} [EM] {electromagnetics}
\newacro{era} [ERA] {electromagnetically reconfigurable antenna}
\newacro{aod} [AOD] {angle-of-departure}
\newacro{aoa} [AOA] {angle-of-arrival}
\newacro{sp} [SP] {scatter point}
\newacro{ml} [ML] {maximum likelihood}
\newacro{mse} [MSE] {mean square error}
\newacro{snr} [SNR] {signal-to-noise ratio}
\newacro{rmse} [RMSE] {root mean square error}
\newacro{crb} [CRB] {Cram\'er-Rao bound}
\newacro{bcrb} [BCRB] {Bayesian Cram\'er-Rao bound}
\newacro{peb} [PEB] {position error bound}
\newacro{kld} [KLD] {Kullback-Leibler divergence}
\newacro{siso} [SISO] {single-input-single-output}
\newacro{mimo} [MIMO] {multiple-input multiple-output}
\newacro{mcrb} [MCRB] {misspecified Cram\'er-Rao bound}
\newacro{bs} [BS] {base station}
\newacro{ue} [UE] {user equipment}
\newacro{arv} [ARV] {array response vector}
\newacro{upa} [UPA] {uniform planar array}
\newacro{rf} [RF] {radio frequency}
\newacro{ris} [RIS] {reconfigurable intelligent surface}
\newacro{rms} [RMS] {root mean square}
\newacro{awgn} [AWGN] {additive white Gaussian noise}
\newacro{nsga2} [NSGA-II] {non‐dominated sorting genetic algorithm II}
\newacro{de} [DE] {differential evolution}
\newacro{iid} [i.i.d.] {independently
and identically distribute}
\newacro{sinr} [SINR] {signal to interference and noise ratio}
\newacro{sdr} [SDR] {semidefinite relaxation}
\newacro{pdf} [PDF] {probability density function}
\newacro{mpc} [MPC] {multipath component}
\newacro{lmi} [LMI] {linear matrix inequality}
\newacro{psd} [PSD] {positive semi-definite}
\newacro{rem} [REM] {radio environment map}
\newacro{it} [IT] {interference temperature}
\newacro{qos} [QoS] {quality-of-service}
\newacro{dnn} [DNN] {deep neural network}
\newacro{drl} [DRL] {deep reinforcement learning}
\newacro{ss} [SS] {spectrum sensing}
\newacro{dft} [DFT] {discrete Fourier transform}
\newacro{ao} [AO] {alternating optimization}
\newacro{se} [SE] {spectral efficiency}
\newacro{mmwave} [mmWave] {millimeter-wave}
\newacro{fpga} [FPGA] {field-programmable gate array}
\newacro{mm_alg} [MM] {majorization-minimization}
\newacro{evd} [EVD] {eigenvalue decomposition}

\newacro{bpm} [BPM] {beampattern matching}
\newacro{mrt} [MRT] {maximum-ratio transmission}
\newacro{rcs} [RCS] {radar cross section}

\newacro{phr} [PHR] {Powell-Hestenes-Rockafellar}

\newacro{gelu} [GELU] {Gaussian Error Linear Unit}
\newacro{relu} [ReLU] {Rectified Linear Unit}
\newacro{oor} [OOR] {Out-of-Region}
\newacro{ann} [ANN] {Artificial Neural Network}
\newacro{mlp} [MLP] {Multi-Layer Perceptron}
\newacro{fnn} [FNN] {Fourier Neural Network}
\newacro{minn} [MINN] {Metasurfaces-Integrated Neural Network}
\newacro{ota} [OTA] {Over-The-Air}
\newacro{pi-fnn} [PI-FNN] {Physics-Informed Fourier Neural Network}

\newacro{disac} [DISAC] {distributed integrated sensing and communication}
\newacro{ap} [AP] {access point}
\newacro{cpu} [CPU] {central processing unit}
\newacro{nr} [NR] {New Radio}
\newacro{veb} [VEB] {velocity error bound}
\newacro{doa} [DOA] {direction of arrival}
\newacro{2d} [2D] {two-dimensional}

\newacro{tx} [TX] {transmitting}
\newacro{rx} [RX] {receiving}
\newacro{sdma} [SDMA] {spatial division multiple access}
\newacro{gbo} [GBO] {gradient-based optimization}

\newacro{ma} [MA] {movable antenna}
\newacro{fpa} [FPA] {fixed-position antenna}

\newacro{fas} [FAS] {fluid antenna system}

\newacro{mm} [MM] {majorization-minimization}
\newacro{zf} [ZF] {zero-forcing}

\newacro{sca} [SCA] {successive convex approximation}

\newacro{gnn} [GNN] {graph neural network}
\newacro{mbgnn}[MBGNN]{model-based graph neural network}
\newacro{ttmbgnn}[TT-MBGNN]{two-timescale model-based graph neural network}
\newacro{smbgnn}[S-MBGNN]{slow-timescale model-based graph neural network}
\newacro{fmbgnn}[F-MBGNN]{fast-timescale model-based graph neural network}

\newacro{rzf}[RZF]{regularized zero-forcing}
\newacro{ste}[STE]{straight-through estimator}
\newacro{silu}[SiLU]{sigmoid linear unit}
\newacro{saa-ao}[SAA-AO]{sample-average-approximation-based alternating optimization}
\newacro{ura}[URA]{uniform rectangular array}
\newacro{adamw}[AdamW]{Adam with decoupled weight decay}

\makeatother
\makeatother
\title{Joint Movable Antenna Positioning and Precoding for ISAC via a Two-Timescale Model-Based GNN}

\author{
Alireza Fadakar, \emph{Graduate Student Member, IEEE}, 
Yuchen Zhang, \emph{Member, IEEE}, 
Andreas F. Molisch, \emph{Fellow, IEEE}\thanks{A. Fadakar and A. F. Molisch are with the Ming Hsieh Department of Electrical and Computer Engineering, University of Southern California, Los Angeles, California, USA; E-mail: \{fadakarg,molisch\}@usc.edu.}
\thanks{Y. Zhang is with the Electrical and Computer Engineering Program, Computer, Electrical and Mathematical Sciences and Engineering (CEMSE), King Abdullah University of Science and Technology (KAUST), Thuwal 23955-6900, Kingdom of Saudi Arabia; E-mail: yuchen.zhang@kaust.edu.sa.}
\thanks{This work is supported 
by the National Science Foundation (Grants 2504811, 2229535 and 2433981). \textit{(Corresponding author: Yuchen Zhang.)}}
}

\usepackage{accents}

\usepackage{amsthm}

\theoremstyle{plain}

\newtheoremstyle{iremark}
  {\topsep}   
  {\topsep}   
  {\upshape}  
  {0.2in}       
  {\itshape}  
  {.}         
  {5pt plus 1pt minus 1pt} 
  {\thmname{#1}\thmnumber{ \itshape#2}\thmnote{ (#3)}} 

\usepackage{amsthm}

\newtheorem{remark}{Remark}

\theoremstyle{definition}
\newtheorem*{proof}{Proof}

\makeatletter
\newcommand*\rel@kern[1]{\kern#1\dimexpr\macc@kerna}
\newcommand*\widebar[1]{%
  \begingroup
  \def\mathaccent##1##2{%
    \rel@kern{0.8}%
    \overline{\rel@kern{-0.8}\macc@nucleus\rel@kern{0.2}}%
    \rel@kern{-0.2}%
  }%
  \macc@depth\@ne
  \let\math@bgroup\@empty \let\math@egroup\macc@set@skewchar
  \mathsurround\z@ \frozen@everymath{\mathgroup\macc@group\relax}%
  \macc@set@skewchar\relax
  \let\mathaccentV\macc@nested@a
  \macc@nested@a\relax111{#1}%
  \endgroup
}
\makeatother

\begin{document}
\maketitle
\begin{abstract}
\Acp{ma} allow an \ac{isac} \ac{bs} to reshape multiuser channels and angular sensitivity.
For monostatic multiuser communication and multi-target elevation and azimuth estimation, we minimize the expected target-averaged angular \ac{bcrb} subject to per-user block-level and long-term rate requirements.
To avoid repeated online optimization, we propose a \ac{ttmbgnn} over antennas, users, and targets.
The slow policy selects positions from long-term statistics to reduce movement overhead, while the fast policy learns complex corrections to a \ac{rzf} reference using post-movement channels and target-angle priors.
We jointly train both policies through a differentiable model of rates and \ac{bcrb}, without optimization-generated labels.
Across tested rate requirements, the network yields lower sensing bounds than variants using a fixed array, \ac{rzf}, or both, with any-user fast-rate violations in $0.31$--$2.63\%$ of test blocks.
Frozen policies provide competitive sensing bounds for unseen user and target counts, and ten fine-tuning epochs improve communication reliability at higher user counts.
Across tested network sizes, our framework achieves up to
$3$~dB lower \ac{bcrb} than an idealized alternating-optimization benchmark that uses known target reflection coefficients.
The network computes precoders approximately $50$-fold faster
in matched batch-one tests.
\end{abstract}

\acresetall


\bstctlcite{IEEEexample:BSTcontrol}

\vspace{-0.5cm}
\section{Introduction}

\Ac{isac} supports communication and environmental sensing through
shared \ac{rf} components, transmitted signals, and spectral
resources~\cite{Liu2022ISAClimits,Lu2023ISACChallenges}.
This integration can improve spectral and hardware efficiency, while
the \ac{bs} array geometry shapes both communication and sensing
performance.
Most existing systems employ \acp{fpa} whose element locations remain
unchanged after deployment.
Digital precoding controls the complex antenna excitations, while
the spatial sampling positions remain set by the deployed geometry.
Consequently, fixed geometries limit the flexibility to reshape
multiuser channel correlations, mitigate interference, and adjust
the effective sensing aperture.

\Ac{ma} technology addresses this limitation by allowing antenna
elements or subarrays to change their positions within prescribed
movement regions~\cite{Zhu2024MAOverview,Ma2026RAMASurvey}.
Closely related position-reconfigurable architectures are also
studied under the \ac{fas}
terminology~\cite{Wong2023FAS,Ma2026RAMASurvey,fadakar2026hybrid}.
The \ac{fas} concept includes liquid conductors, switchable ports,
reconfigurable pixels, and mechanically repositioned antennas.
This paper specifically considers \acp{ma} with continuously
adjustable physical coordinates.
Such reconfiguration can improve channel gains, reduce multiuser
interference, and reshape the sensing response without increasing
the number of \ac{rf} chains.
These benefits have motivated joint position and beamforming
designs for communication
systems~\cite{Zhu2024MAOverview,Zhang2025MAHybrid} and for
\ac{isac} systems based on sensing \ac{sinr}, beampattern, or
\ac{crb} objectives~\cite{Jiang2025MAISAC,Chen2025MAISAC_CRB,
Ma2026MAISAC_StatCSI,Han2025MA_MU_ISAC}.
However, antenna positions enter the communication channels and sensing steering vectors through nonlinear phase terms and are tightly coupled with the precoders.
The resulting nonconvex problems are commonly addressed using
iterative combinations of \ac{ao}, \ac{sca}, \ac{mm}, penalty or
dual-decomposition methods, and gradient-based
updates~\cite{Zhang2025MAHybrid,Jiang2025MAISAC,Chen2025MAISAC_CRB,
Qin2024MAISAC_CRB,Han2025MA_MU_ISAC}.

A further challenge arises from the different adaptation rates of mechanical positioning and electronic precoding.
Moving the antennas in every coherence block can incur excessive delay and positioning
overhead~\cite{Ma2026MAISAC_StatCSI,Khalili2026MAISAC_TTS}.
Moreover, directly measuring instantaneous channels across many
candidate antenna positions entails additional acquisition overhead.
These limitations motivate slow-timescale positioning using
\ac{ue}-location distributions and long-term propagation statistics,
followed by fast precoding using instantaneous \ac{csi} acquired at
the selected
geometry~\cite{Ma2026MAISAC_StatCSI,Zheng2025TTS_MA}.
In our formulation, the slow inputs additionally include
target-prior statistics.
The sensing objective must also account for block-dependent
uncertainty in the target directions.
The \ac{bcrb} incorporates prior information into estimation-error
bounds~\cite{Gill1995VanTrees}.
We use this bound as a common sensing criterion, with the slow
geometry evaluated across block states and the fast precoder
adapted to the current prior.

Existing optimization-based studies address complementary aspects
of joint positioning and beamforming.
The hybrid architecture in~\cite{Zhang2025MAHybrid} jointly
optimizes movable-subarray positions and analog and digital
beamformers.
Slow positioning based on localized user regions and statistical
\ac{csi} is studied in~\cite{Ma2026MAISAC_StatCSI}, while discrete
and continuous two-timescale designs are developed
in~\cite{Zheng2025TTS_MA,Khalili2026MAISAC_TTS}.
A recent cell-free \ac{mimo} design also optimizes antenna
positions and orientations using global statistical \ac{csi},
while updating decentralized beamformers on a faster
timescale~\cite{Zhang2026TwoTimescale6DMA}.
These studies establish the value of coordinating mechanical
positioning and electronic precoding across different timescales.

Existing iterative designs \textit{typically require repeated optimization as operating conditions change}, increasing online computational cost.
Learning-based methods address this limitation by transferring much of the optimization effort to offline
training~\cite{Shen2023GNN,Lu2025GNN_Wireless}.
In particular, \acp{gnn} represent interactions among network
entities through shared message functions and have been applied to
beamforming, power allocation, and energy-efficiency
optimization~\cite{Li2024gnn_sumrate,Zhang2025mbgnn_energy}.
Shared parameters also allow the same graph-layer structure to
accommodate different numbers of users and targets.
Heterogeneous graph architectures for cell-free and cooperative
\ac{isac} generally assume fixed antenna
geometries~\cite{wang2024heterogeneous,Lian2026CoISAC_GNN,
Demirhan2024GNN}.

For \ac{ma} and \ac{fas} systems, learning-based studies have explored
multi-agent and deep reinforcement learning, generative learning,
and hierarchical policies for movement, port selection, and
beamforming~\cite{Weng2024MA_MADDPG,Wang2024FAS_ISAC_DRL,
Zhang2026FAS_GenRL,Sheraz2026FAS_HDRL}.
The graph-based designs
in~\cite{He2025FAS_GNN,Xiu2025MA_CFISAC_GNN} address positioning
and beamforming without separating statistical geometry selection
from block-adaptive precoding, while the two-timescale
deep-unfolding method in~\cite{Liang2026TTS_Deep} considers
communication objectives.
Supervised alternatives based on optimization-generated labels
also incur the cost of repeatedly executing a numerical solver.
A \ac{mbgnn} can embed differentiable physical models into
training~\cite{Zhang2025mbgnn_energy}, enabling direct objective
evaluation without these labels.
A gap therefore remains in
\textit{jointly learning long-term positioning and
geometry-dependent precoding} for Bayesian multi-target angle
estimation under communication requirements at both timescales.

Motivated by these observations, we propose a \ac{ttmbgnn} for
monostatic \ac{ma}-assisted \ac{isac}.
The \ac{smbgnn} selects a common geometry from long-term statistics,
and the \ac{fmbgnn} adapts the precoder to current block information.
Both policies are trained jointly without optimization-generated
labels through a differentiable model that evaluates communication
rates and target-averaged angular \acp{bcrb}.
The $K$ user data streams also support sensing, without dedicated
sensing streams~\cite{Liu2022ISAClimits,Qin2024MAISAC_CRB,
Han2025MA_MU_ISAC}.
Target ranges and reflection coefficients enter only the sensing
model and are excluded from the policy inputs.

\vspace{-0.3cm}

\subsection{Main Contributions}

The main contributions are summarized as follows.
\begin{itemize}
\item
We formulate a two-timescale stochastic design for monostatic \ac{ma}-assisted \ac{isac}. 
One continuous antenna geometry is chosen per slow interval from \ac{ue}-location and channel statistics, target-angle prior statistics, and long-term rate requirements.
Each block precoder adapts to post-movement \ac{csi}, current target priors, and fast-timescale rate requirements. The objective minimizes the expected target-averaged angular \ac{bcrb} subject to per-\ac{ue} block and long-term rate requirements, transmit power, antenna
movement, and minimum-spacing constraints.

\item
We develop a heterogeneous \ac{ttmbgnn} with an \ac{smbgnn} for geometry selection and an \ac{fmbgnn} for block-adaptive precoding.
Relation-specific messages combine attention and mean aggregation over \ac{ma}, \ac{ue}, and target nodes, and complex \ac{ma}-\ac{ue} channels affect the fast policy as edge features. 
The proposed decoder learns entrywise complex corrections to a \ac{rzf} reference and normalizes the resulting precoder to meet the power limit. 
Shared parameters make the graph layers equivariant to consistent node relabeling at inference and support different node counts without changing the layer parameterization.

\item
We train both policies jointly without optimization-generated labels.
A differentiable communication-sensing model evaluates block rates and angular \acp{bcrb}, providing block-specific precoder gradients and a sample-average sensing objective for the common geometry.
One-sided augmented-Lagrangian losses and projected multiplier updates address fast and slow \ac{qos} requirements and raw antenna spacing. 
Structured outputs satisfy the power and movement-region limits.
Although the proposed finite geometry repair does not provide a universal spacing guarantee, every repaired geometry checked in the reported simulations met the minimum antenna spacing requirement.

\item
Simulations on unseen realizations demonstrate lower sensing bounds than learned-component ablations and fixed-array \ac{rzf} across the tested \ac{qos} levels.
Across the evaluated network-size configurations, including fine-tuned policies, the proposed framework achieves up to $3$~dB lower angular \ac{bcrb} than the idealized \ac{saa-ao} benchmark.
Frozen policies provide competitive sensing bounds at unseen user and target counts, while ten additional epochs improve communication reliability at larger user counts.
Batch-one per-block precoder computation is approximately $50$ times faster than iterative refinement in the considered scenario.
\end{itemize}

\paragraph*{Notation}
Scalars, vectors, and matrices are denoted by regular lowercase, boldface lowercase, and boldface uppercase letters, respectively, while calligraphic letters denote sets or mappings as defined in context. 
The superscripts $(\cdot)^{\mathsf T}$ and $(\cdot)^{\mathsf H}$ represent transpose and Hermitian transpose, respectively, and $j=\sqrt{-1}$. 
The operators $\Re\{\cdot\}$, $\Im\{\cdot\}$, and $|\cdot|$ return the real part, imaginary part, and scalar magnitude. The symbols $\lVert\cdot\rVert_2$ and $\lVert\cdot\rVert_{\mathrm F}$ denote the Euclidean and Frobenius norms, respectively. 
The $n\times n$ identity matrix is denoted by $\mathbf I_n$, and $\mathbf 1_n$ is the $n$-dimensional all-ones vector. 
We use $\operatorname{col}(\cdot)$ for vertical stacking of scalar or vector arguments, $\Vert$ for concatenation, and $\odot$ for the Hadamard product.
Finally, $[x]_+=\max\{x,0\}$ denotes the positive-part operator, $|\mathcal S|$ is the cardinality of set $\mathcal S$, and $\Pi_{\mathcal C}(\cdot)$ denotes Euclidean projection onto the set $\mathcal C$.

\vspace{-0.3cm}
\section{System and Signal Model}
\label{sec:system_model}

\begin{figure}[!t]
\centering
\includegraphics[width=\columnwidth]
{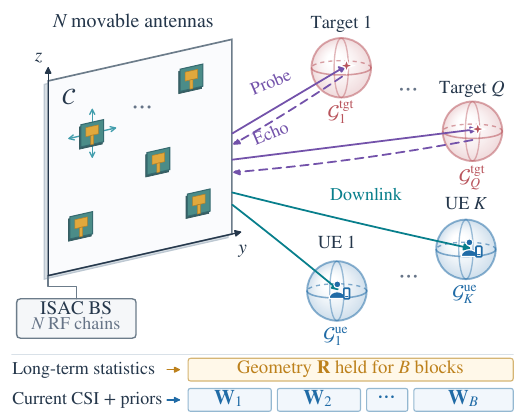}
\caption{Considered monostatic \ac{ma}-assisted \ac{isac} with a
continuous $YZ$-plane aperture.
Shaded spheres denote long-term location regions, and markers
illustrate block-dependent \ac{ue} and nominal target locations.
Teal arrows represent downlink transmission, while solid and
dashed purple arrows denote probing and echoes, respectively.
The common geometry $\mathbf R$ is selected from long-term
statistics and held fixed for $B$ coherence blocks, whereas
the precoder $\mathbf W_b$ is updated per block using current
\ac{csi} and target-angle priors.}
\label{fig:sys_model}
\end{figure}

As illustrated in Fig.~\ref{fig:sys_model}, we consider a monostatic \ac{isac} system in which a dual-functional \ac{bs}, equipped with $N$ continuously adjustable \acp{ma} and $N$ dedicated \ac{rf} chains, serves $K$ single-antenna \acp{ue} while supporting \ac{2d} \ac{doa} estimation for $Q$ point targets in elevation and azimuth\footnote{Array-based angular estimation exploits spatial phase
differences under the adopted narrowband, frequency-flat
model~\cite{Ma2026MAISAC_StatCSI}. Range and Doppler estimation are
outside the present scope.}.
The corresponding index sets are
$\mathcal N\triangleq\{1,\ldots,N\}$,
$\mathcal K\triangleq\{1,\ldots,K\}$, and
$\mathcal Q\triangleq\{1,\ldots,Q\}$.
We adopt a two-timescale positioning and precoding design in which the antenna geometry is selected using long-term statistical information and held fixed over a positioning interval of $B$ transmission blocks, indexed by
$\mathcal B\triangleq\{1,\ldots,B\}$.
Each block contains $T$ symbol intervals, indexed by $\mathcal T\triangleq\{1,\ldots,T\}$.
The digital precoder is updated in each block using the selected geometry and current block information. 
The statistical inputs remain valid throughout the positioning interval, whereas the instantaneous conditions are approximated as constant within each block and may vary between blocks.
To describe spatial variability, we associate \ac{ue} $k$ and target $q$
with spherical location regions
\begin{align}
\mathcal G_k^{\mathrm{ue}}
&\triangleq
\left\{\mathbf p\in\mathbb R^3:
\|\mathbf p-\overline{\mathbf p}_k^{\mathrm{ue}}\|_2
\leq r_k^{\mathrm{ue}}\right\},
\label{eq:ue_location_region}\\
\mathcal G_q^{\mathrm{tgt}}
&\triangleq
\left\{\mathbf p\in\mathbb R^3:
\|\mathbf p-\overline{\mathbf p}_q^{\mathrm{tgt}}\|_2
\leq r_q^{\mathrm{tgt}}\right\},
\label{eq:target_location_region}
\end{align}
where $\overline{\mathbf p}_k^{\mathrm{ue}},
\overline{\mathbf p}_q^{\mathrm{tgt}}\in\mathbb R^3$ are the region
centers and $r_k^{\mathrm{ue}},r_q^{\mathrm{tgt}}\geq0$ are their
radii. These parameters specify the regional distributions, which
remain unchanged within each positioning interval. In block $b$, the
\ac{ue} and nominal target locations are sampled independently across
users, targets, and blocks as
\begin{equation}
\begin{aligned}
\mathbf p_{k,b}^{\mathrm{ue}}
\sim\mathcal U_{\mathrm{vol}}(\mathcal G_k^{\mathrm{ue}}),
\mathbf p_{q,b}^{\mathrm{tgt}}
\sim\mathcal U_{\mathrm{vol}}(\mathcal G_q^{\mathrm{tgt}}),
\end{aligned}
\label{eq:block_location_sampling}
\end{equation}
where $\mathcal U_{\mathrm{vol}}$ denotes the uniform distribution with respect to volume over the specified region. The sampled locations remain unchanged within a block. 
The nominal target location determines
the center of the block-dependent angular prior introduced in Section~\ref{subsec:sensing_bcrb}.
Residual transmitter leakage and hardware-induced self-interference are assumed sufficiently suppressed and are omitted from the signal model~\cite{Ma2026MAISAC_StatCSI}.

\begin{remark}[Statistical inputs and temporal information]
As detailed in Section~\ref{sec:proposed_hmbgnn}, we provide the \ac{fmbgnn} with current block information and the \ac{smbgnn} with long-term statistical summaries available before antenna positioning.
These summaries may be estimated from historical observations.
Neither network explicitly processes an ordered block sequence or maintains a recurrent state. 
Under the stationary model, we use independent snapshots to estimate the statistical summaries and expected training loss. 
We leave explicit trajectory modeling and
history-dependent policy learning for future work.
\end{remark}

\vspace{-0.5cm}
\subsection{Movable-Antenna Array Architecture}
\label{subsec:ma_architecture}
A local Cartesian coordinate system is considered at the \ac{bs}, with its boresight aligned with the positive $X$-axis and the continuous antenna-movement region $\mathcal{C}\subset\mathbb{R}^{2}$ located on the $YZ$-plane. 
The position of the $n$th \ac{ma} is $\mathbf{r}_{n}=[y_n,z_n]^{\mathsf T}\in\mathcal{C}$, with the corresponding three-dimensional coordinate $\bar{\mathbf{r}}_{n}=[0,y_n,z_n]^{\mathsf T}$. The complete array geometry is collected as $\mathbf{R}\triangleq[\mathbf{r}_{1},\ldots,\mathbf{r}_{N}]\in\mathbb{R}^{2\times N}$. The movement-region boundaries, minimum inter-antenna separation, and other physical constraints are specified in the optimization formulation.
Let
$\boldsymbol{\vartheta}
=[\vartheta^{\mathrm{el}},\vartheta^{\mathrm{az}}]^{\mathsf T}$
denote a far-field \ac{2d} \ac{doa}, where
$\vartheta^{\mathrm{el}}\in[-\pi/2,\pi/2]$ is measured from the $XY$-plane toward the positive $Z$-axis and $\vartheta^{\mathrm{az}}\in[-\pi,\pi)$ is measured in the $XY$-plane from the positive $X$-axis toward the positive $Y$-axis. 
The corresponding unit direction vector is defined as
$\mathbf{u}(\boldsymbol{\vartheta})
=[\cos(\vartheta^{\mathrm{el}})\cos(\vartheta^{\mathrm{az}}),
\cos(\vartheta^{\mathrm{el}})\sin(\vartheta^{\mathrm{az}}),
\sin(\vartheta^{\mathrm{el}})]^{\mathsf T}$.
Because the antennas move only on the $YZ$-plane, define the projected direction vector as
$\bar{\mathbf{u}}(\boldsymbol{\vartheta})
=[\cos(\vartheta^{\mathrm{el}})\sin(\vartheta^{\mathrm{az}}),
\sin(\vartheta^{\mathrm{el}})]^{\mathsf T}$. Hence,
$\mathbf{u}^{\mathsf T}(\boldsymbol{\vartheta})
\bar{\mathbf{r}}_{n}
=
\bar{\mathbf{u}}^{\mathsf T}(\boldsymbol{\vartheta})\mathbf{r}_{n}$.
Under the far-field plane-wave model, the position-dependent steering vector is
\begin{equation}
\mathbf{a}
\left(
\mathbf{R},
\boldsymbol{\vartheta}
\right)
=
\exp
\left(
j\frac{2\pi}{\lambda}
\mathbf{R}^{\mathsf T}
\bar{\mathbf{u}}
\left(
\boldsymbol{\vartheta}
\right)
\right)
\in\mathbb{C}^{N},
\label{eq:ma_steering}
\end{equation}
where $\lambda$ is the carrier wavelength and the exponential is applied element-wise. Unlike a conventional fixed array, the manifold in \eqref{eq:ma_steering} is directly controlled by $\mathbf{R}$.
Consequently, moving the antennas modifies both the spatial correlations of the communication channels and the effective aperture and angular sensitivity available for target sensing.

\vspace{-0.5cm}
\subsection{Downlink Communication Signal Model}
\label{subsec:downlink_communication}

\subsubsection{Position-Dependent Geometric Rician Channel}

For the downlink to \ac{ue} $k$, located at $\mathbf p_{k,b}^{\mathrm{ue}}\in\mathcal G_k^{\mathrm{ue}}$ in block $b$, we adopt a narrowband far-field geometric Rician model for continuous \ac{ma} coordinates
\cite{Zhu2024MAChannel,Chen2025MAISAC_CRB,Ma2026MAISAC_StatCSI}.
Let $\beta_{k,b}>0$ denote the large-scale power gain, $d_{k,b}$ the distance from the \ac{bs} reference point to the \ac{ue}, and
$\boldsymbol\vartheta_{k,b}^{\mathrm{LoS}}\in\mathbb R^2$ the \ac{los} departure direction. The corresponding propagation phase is $\phi_{k,b}=2\pi d_{k,b}/\lambda$, and the steering vector is
$\mathbf a_{k,b}^{\mathrm{LoS}}
=\mathbf a(\mathbf R,\boldsymbol\vartheta_{k,b}^{\mathrm{LoS}})\in\mathbb C^N$.
The scattered field is approximated by $L_{\mathrm N}$ effective \ac{nlos} clusters. 
For cluster $\ell$, let
$\boldsymbol\vartheta_{k,b,\ell}^{\mathrm{NLoS}}\in\mathbb R^2$ denote the nominal departure direction and
$g_{k,b,\ell}\sim\mathcal{CN}(0,1)$ the aggregate gain of unresolved diffuse contributions. Independent gains are assumed across clusters\footnote{Equal average powers are adopted for the diffuse components to simplify the channel parameterization~\cite{Zhu2024MAChannel}.
Instantaneous powers remain random, whereas practical cluster mean powers generally differ.}.
Each cluster is represented by the nominal steering vector
$\mathbf a_{k,b,\ell}^{\mathrm{NLoS}}
=\mathbf a(\mathbf R,\boldsymbol\vartheta_{k,b,\ell}^{\mathrm{NLoS}})
\in\mathbb C^N$. The channel vector
$\mathbf h_{k,b}\in\mathbb C^N$ is
\begin{equation}
\begin{aligned}
\mathbf h_{k,b}^{\mathsf H}
={}&\sqrt{\beta_{k,b}}\Bigg(
\sqrt{\tfrac{\kappa_{k,b}}{\kappa_{k,b}+1}}
e^{-j\phi_{k,b}}
(\mathbf a_{k,b}^{\mathrm{LoS}})^{\mathsf T}\\
&+\sqrt{\tfrac{1}{(\kappa_{k,b}+1)L_{\mathrm N}}}
\sum_{\ell=1}^{L_{\mathrm N}}g_{k,b,\ell}
(\mathbf a_{k,b,\ell}^{\mathrm{NLoS}})^{\mathsf T}
\Bigg),
\end{aligned}
\label{eq:geometric_rician_comms_channel}
\end{equation}
where $\kappa_{k,b}\geq0$ is the Rician factor in linear units, drawn from the prescribed long-term distribution.
For $\nu\in\{\mathrm{el},\mathrm{az}\}$, the angular-offset vector
$\boldsymbol\epsilon_{k,b}^{\nu}\in\mathbb R^{L_{\mathrm N}}$ is drawn from
$\mathcal N(\mathbf0,(\sigma_{\mathrm N}^{\nu})^2
\mathbf I_{L_{\mathrm N}})$,
independently between elevation and azimuth. 
Here, $\sigma_{\mathrm N}^{\nu}>0$ is the offset standard deviation in radians, and each offset has support on $\mathbb R$.
The $\ell$th offset is added to $\vartheta_{k,b}^{\mathrm{LoS},\nu}$ to generate the corresponding cluster direction. 
The resulting azimuth is wrapped modulo $2\pi$ onto the adopted azimuth interval, while elevation is restricted to the valid domain of the elevation convention. Thus, the Gaussian law specifies the offsets before these domain mappings.
Since the steering vectors are explicitly parameterized by the antenna coordinate matrix $\mathbf{R}$, any modification to the \ac{ma} geometry dynamically alters the phase of every geometric path. Ultimately, upon establishing $\mathbf{R}$, the \ac{bs} acquires the aggregate block-level channel matrix
\begin{equation}
\mathbf{H}_b = [\mathbf{h}_{1,b},\ldots,\mathbf{h}_{K,b}] \in \mathbb{C}^{N\times K}.
\label{eq:block_channel_matrix}
\end{equation}

\subsubsection{Transmitted and Received Signals}

The \acp{ue} are served simultaneously using linear \ac{sdma}.
Let $\mathbf w_{k,b}\in\mathbb C^N$ denote the precoder for \ac{ue} $k$ and
$\mathbf W_b=[\mathbf w_{1,b},\ldots,\mathbf w_{K,b}]
\in\mathbb C^{N\times K}$ the block precoding matrix.
The independent Gaussian data symbols are collected in
$\mathbf d_b[t]\sim\mathcal{CN}(\mathbf0,\mathbf I_K)$, with independent draws across symbol intervals. The transmitted signal is
\begin{equation}
\mathbf x_b[t]=\mathbf W_b\mathbf d_b[t]\in\mathbb C^N.
\label{eq:comms_tx_signal}
\end{equation}
Conditioned on $\mathbf W_b$, the signal covariance and average transmit-power constraint in each block are
\begin{equation}
\begin{aligned}
\mathbf Q_{x,b}
&\triangleq\mathbb E\!\left[
\mathbf x_b[t]\mathbf x_b^{\mathsf H}[t]
\mid\mathbf W_b\right]
=\mathbf W_b\mathbf W_b^{\mathsf H},\\
\operatorname{tr}(\mathbf Q_{x,b})
&=\|\mathbf W_b\|_{\mathrm F}^2\leq P_{\max}.
\end{aligned}
\label{eq:total_power_constraint}
\end{equation}
The same data-bearing signal is used for sensing, without additional
dedicated sensing streams~\cite{Qin2024MAISAC_CRB,Han2025MA_MU_ISAC}.
The received signal at \ac{ue} $k$ is
\begin{equation}
y_{k,b}[t]
=\mathbf h_{k,b}^{\mathsf H}\mathbf W_b\mathbf d_b[t]
+z_{k,b}[t],
\label{eq:ue_received_signal}
\end{equation}
where $z_{k,b}[t]\sim\mathcal{CN}(0,\sigma_{c,k}^2)$ is additive
noise independent of the transmitted data.

\subsubsection{Instantaneous and Long-Term Rate Metrics}

With Gaussian signaling and the other users' streams treated as
noise, the block-level \ac{sinr} and spectral efficiency of
\ac{ue} $k$ are
\begin{equation}
\gamma_{k,b}
=\frac{|\mathbf h_{k,b}^{\mathsf H}\mathbf w_{k,b}|^2}
{\sum_{i\ne k}|\mathbf h_{k,b}^{\mathsf H}\mathbf w_{i,b}|^2
+\sigma_{c,k}^2},
\quad
R_{k,b}^{\mathrm{inst}}=\log_2(1+\gamma_{k,b}).
\label{eq:ue_instantaneous_rate}
\end{equation}
The rate is measured in bit/s/Hz. Both quantities remain constant
within a block under the assumed channel and precoder model.
To describe block-to-block randomness, define
\begin{equation}
\boldsymbol\Xi_b
\triangleq(\boldsymbol\Xi_b^{\mathrm c},
\boldsymbol\Xi_b^{\mathrm s}),\ 
\boldsymbol\Xi_b^{\mathrm c}
\triangleq\left\{
\mathbf p_{k,b}^{\mathrm{ue}},\kappa_{k,b},\mathbf g_{k,b},
\boldsymbol\epsilon_{k,b}^{\mathrm{el}},
\boldsymbol\epsilon_{k,b}^{\mathrm{az}}
\right\}_{k=1}^K.
\label{eq:communication_block_state}
\end{equation}
Here, $\mathbf g_{k,b}\in\mathbb C^{L_{\mathrm N}}$ collects
the aggregate diffuse cluster gains, and
$\boldsymbol\epsilon_{k,b}^{\nu}\in\mathbb R^{L_{\mathrm N}}$
collects the angular offsets defined above. The sensing state
$\boldsymbol\Xi_b^{\mathrm s}$ and the resulting target-prior
information are specified in
Section~\ref{subsubsec:block_target_priors}.
The location distribution is given in
\eqref{eq:block_location_sampling}, while the gain and angular-offset laws are specified in the preceding channel subsection.
The Rician factor is generated as
\begin{equation}
\begin{aligned}
\kappa_{k,b}^{\mathrm{dB}}
\sim\mathcal U(\kappa_{\min}^{\mathrm{dB}},
\kappa_{\max}^{\mathrm{dB}}),\quad
\kappa_{k,b}&=10^{\kappa_{k,b}^{\mathrm{dB}}/10}.
\end{aligned}
\label{eq:communication_state_distributions}
\end{equation}
The limits and angular spreads are listed in
Table~\ref{tab:simulation_parameters}. Independent draws are used for
the random communication inputs across users and blocks in the
simulations. The generative state is used for sampling and performance
evaluation. During deployment, the fast policy receives the resulting
post-movement \ac{csi} and current target-prior information, without
requiring the individual cluster gains or angular offsets.
For an integrable function $f(\boldsymbol\Xi)$, the Monte Carlo
estimator is obtained as:
\begin{equation}
\begin{aligned}
\widehat{\mathbb E}_M[f(\boldsymbol\Xi)]
\triangleq\frac1M\sum_{m=1}^M f(\boldsymbol\xi^{(m)}),\ 
\boldsymbol\xi^{(m)}
\overset{\mathrm{i.i.d.}}{\sim}
p_{\boldsymbol\Xi},
\end{aligned}
\label{eq:general_sample_average}
\end{equation}
where $m=1,\ldots,M$ indexes independent numerical samples and $b$ indexes physical transmission blocks. The realized rate averaged over one
positioning interval is
$R_k^{\mathrm{SI},(B)}\triangleq
B^{-1}\sum_{b=1}^B R_{k,b}^{\mathrm{inst}}$.
For a given geometry and a fixed rule for block-adaptive precoding, a stationary block-state process yields the ensemble-average rate and its numerical estimate as
\begin{equation}
\begin{aligned}
\overline R_k
&\triangleq\mathbb E[R_k^{\mathrm{SI},(B)}]
=\mathbb E_{\boldsymbol\Xi}[R_k^{\mathrm{inst}}(\boldsymbol\Xi)],\\
\widehat{\overline R}_{k,M}
&\triangleq\widehat{\mathbb E}_M
[R_k^{\mathrm{inst}}(\boldsymbol\Xi)].
\end{aligned}
\label{eq:ue_long_term_rate}
\end{equation}
The equality follows from the common marginal distribution and does not require independent physical blocks. The slow-timescale \ac{qos} criterion constrains this ensemble mean, without guaranteeing
the realized average in every finite interval. 
Thus, $B$ specifies the geometry persistence in physical blocks, whereas $M$ specifies the numerical sample count.

\vspace{-0.5cm}
\subsection{Monostatic Multi-Target Sensing Model}
\label{subsec:sensing_model}
Following representative \ac{ma}-assisted \ac{isac} models~\cite{Qin2024MAISAC_CRB,Ma2026MAISAC_StatCSI,
Khalili2026MAISAC_TTS}, each target is modeled as a far-field point scatterer with a \ac{los} round-trip path. 
Clutter, target-induced multipath, external interference, and residual transmitter leakage are neglected as previously assumed. 
The targets are stationary within each coherence block, such that their angles and reflection coefficients remain constant over its $T$ snapshots but may vary across blocks. Doppler is therefore ignored, and block-dependent angular uncertainty intervals are assumed available from a preceding sensing or tracking stage.

Let
$\boldsymbol{\vartheta}_{q,b}
=[\vartheta_{q,b}^{\mathrm{el}},
\vartheta_{q,b}^{\mathrm{az}}]^{\mathsf T}$
and
$\mathbf{a}_{q,b}
\triangleq
\mathbf{a}(\mathbf{R},\boldsymbol{\vartheta}_{q,b})$
denote the direction and position-dependent steering vector of target
$q$ in block $b$, respectively. 
Since the same array is used for transmission and reception, we obtain the monostatic spatial response as
$\mathbf{A}_{q,b}
\triangleq
\mathbf{a}_{q,b}\mathbf{a}_{q,b}^{\mathsf T}$.
%
Let $\alpha_{q,b}\in\mathbb{C}$ denote the unknown complex reflection coefficient, which absorbs the round-trip path loss, propagation delay, common phase, and \ac{rcs}. 
It is treated as a deterministic nuisance parameter within each block. 
Defining
$\boldsymbol{\vartheta}_{b}
=[\boldsymbol{\vartheta}_{1,b}^{\mathsf T},\ldots,
\boldsymbol{\vartheta}_{Q,b}^{\mathsf T}]^{\mathsf T}$
and
$\boldsymbol{\alpha}_{b}
=[\alpha_{1,b},\ldots,\alpha_{Q,b}]^{\mathsf T}$,
the aggregate sensing channel is:
\begin{equation}
\mathbf{H}_{\mathrm{s},b}
\left(
\mathbf{R},
\boldsymbol{\vartheta}_{b},
\boldsymbol{\alpha}_{b}
\right)
=
\sum_{q=1}^{Q}
\alpha_{q,b}\mathbf{A}_{q,b}.
\label{eq:aggregate_sensing_channel}
\end{equation}
The known data-bearing signal in \eqref{eq:comms_tx_signal} is reused as the sensing probe. 
The echo received in symbol interval $t$ is obtained as
\begin{equation}
\begin{aligned}
\mathbf{y}_{\mathrm{s},b}[t]
&=
\mathbf{H}_{\mathrm{s},b}
\mathbf{x}_{b}[t]
+
\mathbf{z}_{\mathrm{s},b}[t]
\\
&=
\sum_{q=1}^{Q}
\alpha_{q,b}\mathbf{A}_{q,b}
\mathbf{W}_{b}\mathbf{d}_{b}[t]
+
\mathbf{z}_{\mathrm{s},b}[t],
\end{aligned}
\label{eq:sensing_received_signal}
\end{equation}
where
$\mathbf{z}_{\mathrm{s},b}[t]
\sim\mathcal{CN}
(\mathbf{0},\sigma_{\mathrm{s}}^{2}\mathbf{I}_{N})$
is independent across blocks and symbol intervals.
Define
$\mathbf{D}_{b}
=[\mathbf{d}_{b}[1],\ldots,\mathbf{d}_{b}[T]]$ and
$\mathbf{X}_{b}
=[\mathbf{x}_{b}[1],\ldots,\mathbf{x}_{b}[T]]
=\mathbf{W}_{b}\mathbf{D}_{b}$.
Stacking the $T$ sensing snapshots gives
\begin{equation}
\mathbf{Y}_{\mathrm{s},b}
=
\mathbf{H}_{\mathrm{s},b}
\mathbf{X}_{b}
+
\mathbf{Z}_{\mathrm{s},b},
\label{eq:sensing_received_matrix}
\end{equation}
where
$\operatorname{vec}(\mathbf{Z}_{\mathrm{s},b})
\sim
\mathcal{CN}
(\mathbf{0},\sigma_{\mathrm{s}}^{2}\mathbf{I}_{NT})$.

\vspace{-0.5cm}
\subsection{FIM and BCRB Analysis for 2D-DOA Estimation}
\label{subsec:sensing_bcrb}

We construct a separate angular \ac{bcrb} for each block using a hybrid information formulation with random target angles and deterministic
nuisance reflection coefficients~\cite{Bay2008HybridCRB}. 
The standard
regularity and local-unbiasedness conditions for the deterministic-parameter
directions are assumed.
We define the block-level parameter vectors as 
$\boldsymbol{\rho}_b = \operatorname{col}\!\left( \left\{ \Re\{\alpha_{q,b}\}, \Im\{\alpha_{q,b}\} \right\}_{q=1}^Q \right) \in \mathbb{R}^{2Q}$ 
and
$\boldsymbol{\eta}_b = \operatorname{col}\!\left( \boldsymbol{\vartheta}_b, \boldsymbol{\rho}_b \right) \in \mathbb{R}^{4Q}$, 
where $\boldsymbol{\vartheta}_b \in \mathbb{R}^{2Q}$ stacks the elevation and azimuth angles of all $Q$ targets. 
Letting $\mathbf{G}_{i,b} \triangleq \partial\mathbf{H}_{\mathrm{s},b} / \partial\eta_{i,b}$ denote the sensing-channel derivative with respect to the $i$-th entry of $\boldsymbol{\eta}_b$, the block-level data \ac{fim}, accumulated over $T$ transmitted symbols and averaged over the unit-power data symbols, can be obtained via the Slepian-Bangs formula \cite{fadakar2026hybrid, fadakar2025mutual} as
\begin{equation}
\left[ \mathbf{J}_{\mathrm{D},b} \right]_{i,j} = \frac{2T}{\sigma_{\mathrm{s}}^2} \Re \left\{ \operatorname{tr} \left( \mathbf{Q}_{x,b} \mathbf{G}_{i,b}^{\mathsf{H}} \mathbf{G}_{j,b} \right) \right\}.
\label{eq:block_sensing_fim}
\end{equation}
Consequently, $\mathbf{J}_{\mathrm{D},b}$ depends on the fast-timescale precoder through $\mathbf{Q}_{x,b}$ and on the slow-timescale \ac{ma} geometry through the target steering responses and their angular derivatives.

\subsubsection{Block-Dependent Target-Angle Priors}
\label{subsubsec:block_target_priors}

We distinguish the block-to-block distribution of the target prior from the conditional distribution of the unknown target angles within a given block. The former generates diverse target contexts for offline training, while the latter governs the Bayesian sensing model. 
For the nominal target location sampled in
\eqref{eq:block_location_sampling}, the angular-prior center is given by:
\begin{equation}
\mu_{q,b}^{\nu}
=
\vartheta^{\nu}
\!\left(\mathbf p_{q,b}^{\mathrm{tgt}}\right),
\ 
\nu\in\{\mathrm{el},\mathrm{az}\},
\label{eq:block_target_prior_center}
\end{equation}
where $\vartheta^\nu(\cdot)$ maps a Cartesian location to its elevation or
azimuth direction from the \ac{bs}. 
We collect the sensing variables used for block generation and bound
evaluation as
\begin{equation}
\boldsymbol\Xi_b^{\mathrm s}
\triangleq
\left\{
\mathbf p_{q,b}^{\mathrm{tgt}},
\Delta_{q,b}^{\mathrm{el}},
\Delta_{q,b}^{\mathrm{az}},
\alpha_{q,b}
\right\}_{q=1}^{Q}.
\label{eq:sensing_block_state}
\end{equation}
The reflection coefficients enter the outer block-state average but
are excluded from both policy inputs.
The prior information available at the beginning of block $b$ is collected
as
\begin{equation}
\boldsymbol{\zeta}_b
\triangleq
\operatorname{col}
\left(
\left\{
\mu_{q,b}^{\mathrm{el}},
\mu_{q,b}^{\mathrm{az}},
\Delta_{q,b}^{\mathrm{el}},
\Delta_{q,b}^{\mathrm{az}}
\right\}_{q=1}^{Q}
\right)
\in\mathbb R^{4Q}.
\label{eq:block_target_prior_information}
\end{equation}
Thus, the across-block distribution of $\boldsymbol{\zeta}_b$ is induced
by the Cartesian target-location distributions through the nonlinear
angle mapping in \eqref{eq:block_target_prior_center}; the prior centers
are not assumed to be uniformly distributed in angle. In the considered
implementation, each target has configured support widths that remain
constant throughout the slow interval, i.e.,
$\Delta_{q,b}^{\nu}=\Delta_q^\nu$. Hence, only the prior centers vary
across blocks.
Conditional on $\boldsymbol{\zeta}_b$, the local uncertainty around each prior center is modeled by a raised-cosine density. 
This compactly supported prior approaches zero smoothly at its boundaries and satisfies the regularity conditions of the Van Trees inequality
\cite{vanTrees2001Detection,Gill1995VanTrees}. 
Defining
$
\mathcal I_{q,b}^{\nu}
=
\left[
\mu_{q,b}^{\nu}-\tfrac{\Delta_{q,b}^{\nu}}{2},
\,
\mu_{q,b}^{\nu}+\tfrac{\Delta_{q,b}^{\nu}}{2}
\right]
$,
the conditional density of angular component
$\vartheta_{q,b}^{\nu}$ is
\begin{equation}
p_{q,b}^{\nu}
\!\left(
\psi\mid\boldsymbol{\zeta}_b
\right)
=
\tfrac{
1+
\cos\!\left(
\frac{2\pi(\psi-\mu_{q,b}^{\nu})}
     {\Delta_{q,b}^{\nu}}
\right)
}{
\Delta_{q,b}^{\nu}
}
\mathbb I
\!\left\{
\left|\psi-\mu_{q,b}^{\nu}\right|
\leq
\tfrac{\Delta_{q,b}^{\nu}}{2}
\right\}.
\label{eq:block_raised_cosine_angle_prior}
\end{equation}
Assuming conditional independence among targets and angular components,
the joint within-block prior is obtained as
$
p_b
\!\left(
\boldsymbol{\vartheta}_b
\mid
\boldsymbol{\zeta}_b
\right)
=
\prod_{q=1}^{Q}
\prod_{\nu\in\{\mathrm{el},\mathrm{az}\}}
p_{q,b}^{\nu}
\!\left(
\vartheta_{q,b}^{\nu}
\mid
\boldsymbol{\zeta}_b
\right)
$. 
The resulting prior \ac{fim} is obtained as:
\begin{equation}
\mathbf J_{\mathrm P,b}
=
\operatorname{diag}
\left(
\left\{
\tfrac{4\pi^2}
     {(\Delta_{q,b}^{\nu})^2}
\right\}_{q=1,\ldots,Q,\,
         \nu\in\{\mathrm{el},\mathrm{az}\}}
\right)
\in\mathbb R^{2Q\times2Q},
\label{eq:block_angle_prior_fim}
\end{equation}
with diagonal entries ordered consistently with
$\boldsymbol{\vartheta}_b$.

\subsubsection{Fast-Timescale Per-Block BCRB}

For a fixed geometry $\mathbf{R}$, precoder $\mathbf{W}_b$, and prior parameter set $\boldsymbol{\zeta}_b$, the data \ac{fim} must be averaged over the joint target-angle prior corresponding to block $b$. Because the resulting $2Q$-dimensional expectation lacks a tractable closed form, it is approximated using $S_{\mathrm{a}}$ joint angle samples:
\begin{equation}
\overline{\mathbf{J}}_{\mathrm{D},b} = \mathbb{E}_{\boldsymbol{\vartheta}_b \sim p_b} \big[ \mathbf{J}_{\mathrm{D},b} ( \mathbf{R}, \mathbf{W}_b, \boldsymbol{\eta}_b ) \big] 
\approx 
\tfrac{
\sum_{s=1}^{S_{\mathrm{a}}} \mathbf{J}_{\mathrm{D},b} \big( \mathbf{R}, \mathbf{W}_b, \boldsymbol{\eta}_b^{(s)} \big)
}{S_{\mathrm{a}}} ,
\label{eq:block_prior_averaged_data_fim}
\end{equation}
where $\boldsymbol{\eta}_b^{(s)} = \operatorname{col}( \boldsymbol{\vartheta}_b^{(s)}, \boldsymbol{\rho}_b ) \in \mathbb{R}^{4Q}$. The deterministic nuisance parameter vector $\boldsymbol{\rho}_b$ remains fixed across all $s$ samples. To provide a differentiable mapping from the interval parameters to the sampled angles, each sampled component is generated via the affine reparameterization $\vartheta_{q,b}^{\nu,(s)} = \mu_{q,b}^\nu + \Delta_{q,b}^\nu v_{q,\nu}^{(s)}$, where $v_{q,\nu}^{(s)} \in [-1/2, 1/2]$ is drawn from the standardized density $1 + \cos(2\pi v)$. 
Partitioning $\overline{\mathbf J}_{\mathrm D,b}$ into angular
and nuisance blocks, we obtain the equivalent Bayesian angular
information matrix through the Schur complement,
$\mathbf J_{\mathrm B,b}^{\mathrm{eq}}
=\overline{\mathbf J}_{\boldsymbol\vartheta\boldsymbol\vartheta,b}
+\mathbf J_{\mathrm P,b}
-\overline{\mathbf J}_{\boldsymbol\vartheta\boldsymbol\rho,b}
\overline{\mathbf J}_{\boldsymbol\rho\boldsymbol\rho,b}^{-1}
\overline{\mathbf J}_{\boldsymbol\rho\boldsymbol\vartheta,b}$,
assuming that
$\overline{\mathbf J}_{\boldsymbol\rho\boldsymbol\rho,b}$
is nonsingular.
This construction accounts for the unknown reflection coefficients
under the per-block nuisance model in
Section~\ref{subsec:sensing_model}.
The per-block \ac{bcrb} matrix $\mathbf{C}_{\mathrm{BCRB},b} \in \mathbb{R}^{2Q \times 2Q}$ and the target-averaged sensing metric are defined as:
\begin{equation}
\mathbf{C}_{\mathrm{BCRB},b} = \left( \mathbf{J}_{\mathrm{B},b}^{\mathrm{eq}} \right)^{-1}, 
\  
\mathcal{B}_b = \frac{1}{Q} \sum_{q=1}^Q \operatorname{tr} \left( \mathbf{C}_{\mathrm{BCRB},b}^{(q)} \right),
\label{eq:fast_timescale_bcrb}
\end{equation}
where $\mathbf{C}_{\mathrm{BCRB},b}^{(q)} \in \mathbb{R}^{2 \times 2}$ denotes the principal submatrix associated with the elevation and azimuth angles of target $q$.

\subsubsection{Slow-Timescale Expected BCRB}
\label{subsubsec:slow_expected_bcrb}

Under stationary block states, the expected average sensing metric
over the $B$ blocks of a slow interval equals the expected per-block
\ac{bcrb}. Using the sample-average operator in
\eqref{eq:general_sample_average}, the slow-timescale objective and
its Monte Carlo approximation are
\begin{equation}
\begin{aligned}
\overline{\mathcal B}_{\mathrm S}
&\triangleq
\mathbb E_{\boldsymbol\Xi}
\!\left[\mathcal B(\boldsymbol\Xi)\right], \ 
\widehat{\overline{\mathcal B}}_{\mathrm S,M}
\triangleq
\widehat{\mathbb E}_M
\!\left[\mathcal B(\boldsymbol\Xi)\right].
\end{aligned}
\label{eq:slow_timescale_expected_bcrb}
\end{equation}
Here, $B$ specifies the duration of the common geometry in coherence
blocks, whereas $M$ is the number of samples used to approximate
the expectation.

\vspace{-0.5cm}
\section{Two-Timescale Optimization Problem Formulation}
\label{sec:problem_formulation}

We jointly design the \ac{ma} geometry and digital precoders according to their different adaptation timescales. 
The geometry is selected from long-term statistics at the beginning of each slow interval, with mechanical repositioning completed before block transmission.
At the beginning of block $b$, a control stage provides the post-movement \ac{csi}
$\mathbf H_b(\mathbf R)\in\mathbb C^{N\times K}$ and the target-prior information $\boldsymbol\zeta_b\in\mathbb R^{4Q}$. An external tracker or predictor supplies the latter using observations available before
block $b$. 
Acquisition and precoder computation are assumed to finish while both inputs remain valid for the ensuing $T$ communication-sensing symbols.
For $\mathrm X\in\{\mathrm F,\mathrm S\}$, define the requirement vector
$\boldsymbol\rho^{\mathrm X}
=[R_1^{\mathrm X,\min},\ldots,R_K^{\mathrm X,\min}]^{\mathsf T}
\in\mathbb R_+^K$.
A fast policy $\pi_{\mathrm f}$ maps the selected geometry, current
channel, target-prior information, and fast requirements to a precoder:
\begin{equation}
\mathbf W_b
=
\pi_{\mathrm f}\!\left(
\mathbf R,\mathbf H_b(\mathbf R),
\boldsymbol\zeta_b,\boldsymbol\rho^{\mathrm F}
\right)
\in\mathbb C^{N\times K}.
\label{eq:block_fast_policy}
\end{equation}
Here, $\pi_{\mathrm f}$ denotes a general block-adaptive mapping,
whose neural-network parameterization is introduced in
Section~\ref{sec:proposed_hmbgnn}.

\vspace{-0.4cm}
\subsection{Fast-Timescale Digital-Precoder Problem}
\label{subsec:fast_problem}

For a fixed feasible geometry $\mathbf R\in\mathbb R^{2\times N}$, we formulate the model-based blockwise precoder-design problem
$\mathcal P_{\mathrm F,b}$ as
\begin{subequations}
\label{eq:fast_optimization_problem}
\begin{align}
\min_{\mathbf W_b\in\mathbb C^{N\times K}}
\quad&
\mathcal B_b(\mathbf W_b;\mathbf R)
\label{eq:fast_optimization_objective}
\\
\mathrm{s.t.}\quad&
R_{k,b}^{\mathrm{inst}}(\mathbf R,\mathbf W_b)
\geq R_k^{\mathrm F,\min},
\quad \forall k\in\mathcal K,
\label{eq:fast_rate_constraint}
\\
&
\|\mathbf W_b\|_{\mathrm F}^2\leq P_{\max}.
\label{eq:fast_power_constraint}
\end{align}
\end{subequations}
The objective is the block-level \ac{bcrb} in
\eqref{eq:fast_timescale_bcrb}, evaluated using $\boldsymbol\zeta_b$
and the block's fixed nuisance realization $\boldsymbol\rho_b$.
Simulated nuisance values are used for offline training and bound
evaluation and are excluded from the learned policy inputs. For compactness, $R_{k,b}^{\mathrm{inst}}(\mathbf R,\mathbf W_b)$ denotes the rate in \eqref{eq:ue_instantaneous_rate} evaluated with $\mathbf h_{k,b}(\mathbf R)$.
The problem is nonconvex because the precoding vectors are coupled through multiuser interference and the sensing information matrix.

\vspace{-0.5cm}
\subsection{Slow-Timescale MA-Position Problem}
\label{subsec:slow_problem}

The common geometry is selected before the future block states are
known. Each candidate $\mathbf R$ is therefore evaluated through
the fast policy $\pi_{\mathrm f}$. For Monte Carlo sample
$m=1,\ldots,M$, let $\mathbf W^{(m)}$ denote the output of
\eqref{eq:block_fast_policy} using $\mathbf H^{(m)}(\mathbf R)$ and
$\boldsymbol\zeta^{(m)}$. Both the channel and the resulting
precoder depend on the candidate geometry, even when the sampled
environment is unchanged. 
Thus, $\mathbf W^{(m)}$ is determined by $\mathbf R$ and $\pi_{\mathrm f}$ and it is not treated as an independent optimization variable.
We write
$\widehat{\overline{\mathcal B}}_{\mathrm S,M}
(\mathbf R,\pi_{\mathrm f})$ and
$\widehat{\overline R}_{k,M}(\mathbf R,\pi_{\mathrm f})$
for the estimators in
\eqref{eq:slow_timescale_expected_bcrb} and
\eqref{eq:ue_long_term_rate}, respectively, evaluated using these
geometry-dependent channels and policy-generated precoders.

Following the continuous-position \ac{ma} model in
\cite{Xiao2024MultiuserMA,Ma2026MAISAC_StatCSI}, define
$\mathcal C=[y_{\min},y_{\max}]\times[z_{\min},z_{\max}]$
as the movement region and $D_{\min}$ as the minimum inter-\ac{ma}
spacing. For a fixed fast policy $\pi_{\mathrm f}$, the
sample-average geometry problem
$\mathcal P_{\mathrm S}(\pi_{\mathrm f})$ is
\begin{subequations}
\label{eq:slow_optimization_problem}
\begin{align}
\min_{\mathbf R\in\mathbb R^{2\times N}}
\quad&
\widehat{\overline{\mathcal B}}_{\mathrm S,M}
(\mathbf R,\pi_{\mathrm f})
\label{eq:slow_optimization_objective}
\\
\mathrm{s.t.}\quad&
\widehat{\overline R}_{k,M}(\mathbf R,\pi_{\mathrm f})
\geq R_k^{\mathrm S,\min},
\quad \forall k\in\mathcal K,
\label{eq:slow_rate_constraint}
\\
&
\mathbf r_n\in\mathcal C,
\quad \forall n\in\mathcal N,
\label{eq:ma_region_constraint}
\\
&
\|\mathbf r_n-\mathbf r_\ell\|_2\geq D_{\min},
\quad 1\leq n<\ell\leq N,
\label{eq:ma_separation_constraint}
\\
&
R_k^{\mathrm{inst},(m)}(\mathbf R,\mathbf W^{(m)})
\geq R_k^{\mathrm F,\min},
\quad \forall k,m,
\label{eq:slow_sample_fast_feasibility}
\\
&
\|\mathbf W^{(m)}\|_{\mathrm F}^2\leq P_{\max},
\quad \forall m.
\label{eq:slow_sample_power_feasibility}
\end{align}
\end{subequations}
The last two constraints enforce the fast-timescale requirements
on the sampled states. 
Satisfaction of the rate requirements on
unseen states is assessed separately in the numerical evaluation.
The thresholds $R_k^{\mathrm F,\min}$ and
$R_k^{\mathrm S,\min}$ specify block-level and long-term service
requirements, respectively, and need not be equal.
Although $\mathcal P_{\mathrm S}(\pi_{\mathrm f})$ expresses the geometry subproblem for a fixed fast policy, the proposed framework jointly trains the slow geometry policy and $\pi_{\mathrm f}$.
Consequently, both policies account for the sensing objective and the fast- and slow-timescale communication requirements.

\vspace{-0.5cm}
\section{Proposed Two-Timescale Model-Based GNN}
\label{sec:proposed_hmbgnn}

\subsection{Architecture and Information Flow}
\label{subsec:tt_mbgnn_overview}

\begin{figure*}[!t]
\centering
\includegraphics[width=\textwidth]
{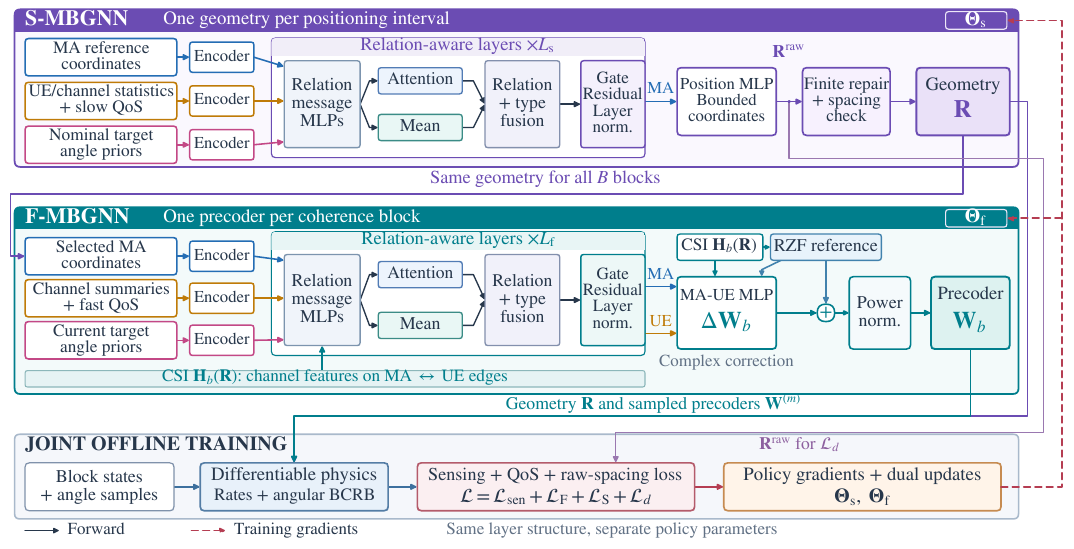}
\caption{Architecture of the proposed \ac{ttmbgnn}.
The \ac{smbgnn} selects one geometry per positioning interval, and the \ac{fmbgnn} produces one precoder per block.
Final \ac{ma} states feed the geometry decoder, while final \ac{ma} and \ac{ue} states feed the precoder decoder.
Target information reaches both decoders through message passing.
Channel and \ac{rzf} coefficient features also enter the pairwise decoder. 
Joint offline training uses the differentiable physics model and raw-spacing loss, with a \ac{ste} for finite repair.
Solid and dashed red arrows denote forward computation and training gradients. 
The policies have the same layer structure and separate parameters.}
\label{fig:tt_mbgnn_architecture}
\vspace{-0.56cm}
\end{figure*}

The proposed \ac{ttmbgnn} separates slow-timescale geometry selection
from block-adaptive precoding while training the two decisions jointly.
As illustrated in Fig.~\ref{fig:tt_mbgnn_architecture}, the
\ac{smbgnn} selects one \ac{ma} geometry per positioning interval from
long-term information. Given this geometry, the \ac{fmbgnn} produces
one precoder in each of the following $B$ coherence blocks using the
current block information. The corresponding policy mappings are represented as:
\begin{align}
\mathbf R^{\mathrm{raw}}
&=
\pi_{\mathrm s}
\!\left(\mathcal X_{\mathrm s};\boldsymbol\Theta_{\mathrm s}\right),
\qquad
\mathbf R
=
\mathcal P_{\mathrm{MA}}\!\left(\mathbf R^{\mathrm{raw}}\right),
\label{eq:slow_gnn_mapping_revised}
\\
\mathbf W_b
&=
\pi_{\mathrm f}\!\left(
\mathbf R,\mathbf H_b(\mathbf R),\boldsymbol\zeta_b,
\boldsymbol\rho^{\mathrm F};
\boldsymbol\Theta_{\mathrm f}\right).
\label{eq:fast_gnn_mapping_revised}
\end{align}
Here, $\mathbf R^{\mathrm{raw}},\mathbf R\in\mathbb R^{2\times N}$
are the raw and repaired geometries, respectively, and
$\mathcal P_{\mathrm{MA}}:\mathbb R^{2\times N}\rightarrow
\mathbb R^{2\times N}$ denotes the finite spacing-repair procedure.
The vectors $\boldsymbol\Theta_{\mathrm s}\in\mathbb R^{P_{\mathrm s}}$
and $\boldsymbol\Theta_{\mathrm f}\in\mathbb R^{P_{\mathrm f}}$
contain the trainable parameters of the two policies. 

\subsubsection{Slow-Timescale Input Features}

The input to the \ac{smbgnn} is collected as
\begin{equation}
\mathcal X_{\mathrm s}
\triangleq
\left(
\mathbf X_{\mathrm s}^{\mathrm{ma}},
\mathbf X_{\mathrm s}^{\mathrm{ue}},
\mathbf X_{\mathrm s}^{\mathrm{tgt}}
\right).
\label{eq:slow_feature_map_compact}
\end{equation}
The matrix
$\mathbf X_{\mathrm s}^{\mathrm{ma}}\in\mathbb R^{N\times2}$
contains the normalized reference coordinates of the $N$ \acp{ma},
where the reference geometry is
$\mathbf R^{\mathrm{ref}}\in\mathbb R^{2\times N}$.
Each row of
$\mathbf X_{\mathrm s}^{\mathrm{ue}}\in\mathbb R^{K\times7}$
describes one \ac{ue} through its long-term departure direction,
large-scale channel gain, Rician factor, and slow-timescale
requirement. Hence, $\boldsymbol\rho^{\mathrm S}$ is already included
in $\mathcal X_{\mathrm s}$. Finally, each row of
$\mathbf X_{\mathrm s}^{\mathrm{tgt}}\in\mathbb R^{Q\times6}$
encodes the nominal elevation and azimuth prior centers and their
angular widths for one target. Please refer to
Appendix~\ref{app:feature_construction} for the explicit slow-feature
definitions and normalization.

\subsubsection{Fast-Timescale Input Features}

For block $b$, the \ac{fmbgnn} input features are collected as
\begin{equation}
\mathcal X_{\mathrm f,b}
\triangleq
\left(
\mathbf X_{\mathrm f,b}^{\mathrm{ma}},
\mathbf X_{\mathrm f,b}^{\mathrm{ue}},
\mathbf X_{\mathrm f,b}^{\mathrm{tgt}},
\mathbf E_b^{\mathrm{ch}}
\right).
\label{eq:fast_feature_map_compact}
\end{equation}
The matrix $\mathbf X_{\mathrm f,b}^{\mathrm{ma}}\in\mathbb R^{N\times2}$ contains the normalized antenna coordinates of the selected geometry $\mathbf R$.
Each row of
$\mathbf X_{\mathrm f,b}^{\mathrm{ue}}\in\mathbb R^{K\times4}$ summarizes the current channel magnitudes and includes the corresponding fast-timescale rate requirement.
The matrix
$\mathbf X_{\mathrm f,b}^{\mathrm{tgt}}\in\mathbb R^{Q\times6}$ encodes the current target-angle prior centers and widths from
$\boldsymbol\zeta_b\in\mathbb R^{4Q}$.
The edge-feature tensor
$\mathbf E_b^{\mathrm{ch}}\in\mathbb R^{N\times K\times3}$ provides three real-valued features for each complex coefficient in $\mathbf H_b(\mathbf R)\in\mathbb C^{N\times K}$.
The same feature vector is assigned to both directed
\ac{ma}-\ac{ue} edges associated with the corresponding link.
Please refer to Appendix~\ref{app:feature_construction} for the explicit feature definitions and normalization.

During training, the differentiable communication-sensing model
evaluates the achieved rates and angular \acp{bcrb} and supplies
gradients to both policies. This model-based procedure requires no
optimization-generated labels~\cite{Eisen2020REGNN}. During deployment,
fixed-depth policy evaluations and output processing replace repeated
optimization of the physical variables.

\vspace{-0.5cm}
\subsection{Heterogeneous Graph and Relation-Aware Layers}
\label{subsec:heterogeneous_graph_revised}

Both policies operate on the directed heterogeneous graph
$\mathcal V=\mathcal V_{\mathrm{ma}}\cup\mathcal V_{\mathrm{ue}}
\cup\mathcal V_{\mathrm{tgt}}$, containing $N$ \ac{ma}, $K$ \ac{ue},
and $Q$ target nodes. 
The five relation types are given by:
\begin{equation}
\begin{aligned}
\mathcal R_{\mathrm g}=\{&
\mathrm{ue}{\to}\mathrm{ma},\,
\mathrm{ma}{\to}\mathrm{ue},\,
\mathrm{tgt}{\to}\mathrm{ma},
\mathrm{ue}{\to}\mathrm{tgt},\,
\mathrm{tgt}{\to}\mathrm{ue}\}.
\end{aligned}
\label{eq:relation_types_revised}
\end{equation}
Each relation is a complete directed bipartite connection between
its source and destination types. For relation $r:s\to t$ and
destination $v\in\mathcal V_t$, the incoming neighborhood is
$\mathcal N_r(v)=\mathcal V_s$. The two \ac{ma}-\ac{ue}
relations also carry channel edge features in the \ac{fmbgnn}.
For policy $y\in\{\mathrm{s},\mathrm{f}\}$, let
$\mathbf x_{y,v}\in\mathbb R^{d_{y,t}}$ be the column-vector feature of node $v\in\mathcal V_t$, with dimension
$d_{y,t}$, from
\eqref{eq:slow_feature_map_compact} or
\eqref{eq:fast_feature_map_compact}. A policy- and type-specific encoder $E_{y,t}:\mathbb R^{d_{y,t}}\to\mathbb R^d$ initializes
$\mathbf z_{y,v}^{(0)}=E_{y,t}(\mathbf x_{y,v})$.
The state $\mathbf z_{y,v}^{(i)}\in\mathbb R^d$ is then updated through $L_y$ relation-aware layers. In the equations below, we fix policy $y$ and, for the fast policy, coherence block $b$, and write $\mathbf z_v^{(i)}\equiv\mathbf z_{y,v}^{(i)}$.
Consequently, $\mathbf z_v^{(i)}$ and $\mathbf z_u^{(i)}$ are
the destination and source states at the same layer: both are
encoder outputs when $i=0$ and outputs of layer $i-1$ otherwise.
Relation-specific message functions distinguish the physical interactions among the node types\cite{Schlichtkrull2018RGCN}. 
Attention permits different source nodes to contribute unequally~\cite{Velickovic2018GAT}, while parallel mean aggregation provides a neighborhood summary normalized by the number of
source nodes. 
The gated residual update subsequently combines these messages without discarding the node's previous state. For $u\in\mathcal N_r(v)$ and
$i=0,\ldots,L_y-1$, the relation-specific \ac{mlp}
$\phi_r^{(i)}:\mathbb R^{2d+d_{e,r}}\to\mathbb R^d$ computes
\begin{equation}
\mathbf m_{u\to v,r}^{(i)}
=\phi_r^{(i)}\!\left(
\mathbf z_v^{(i)}\Vert\mathbf z_u^{(i)}
\Vert\mathbf e_{u\to v,r}\right).
\label{eq:destination_conditioned_message}
\end{equation}
Here, $\Vert$ denotes concatenation. The feature
$\mathbf e_{u\to v,r}\in\mathbb R^{d_{e,r}}$ has
$d_{e,r}=3$ for the two fast \ac{ma}-\ac{ue} relations and
is omitted otherwise ($d_{e,r}=0$).
Let $A$ divide $d$, set $d_h=d/A$, and partition each message
into head vectors
$\mathbf m_{u\to v,r}^{(i,a)}\in\mathbb R^{d_h}$.
For a learned matrix
$\mathbf C_r^{(i)}\in\mathbb R^{A\times d}$, the head score is
$s_{u\to v,r}^{(i,a)}
=[\mathbf C_r^{(i)}\mathbf m_{u\to v,r}^{(i)}]_a/
\sqrt{d_h}$. Normalizing separately over the sources of each
relation gives
\begin{equation}
\alpha_{u\to v,r}^{(i,a)}
=
\frac{\exp(s_{u\to v,r}^{(i,a)})}
{\sum_{u'\in\mathcal N_r(v)}
 \exp(s_{u'\to v,r}^{(i,a)})},
\quad a=1,\ldots,A.
\label{eq:relation_attention_weights}
\end{equation}
The attention and mean aggregates, followed by their
relation-specific fusion, are
\begin{equation}
\begin{aligned}
\mathbf a_{v,r}^{\mathrm{att},(i)}
&=\mathop{\Vert}_{a=1}^{A}
  \sum_{u\in\mathcal N_r(v)}
  \alpha_{u\to v,r}^{(i,a)}
  \mathbf m_{u\to v,r}^{(i,a)},\\
\mathbf a_{v,r}^{\mathrm{mean},(i)}
&=\frac{1}{|\mathcal N_r(v)|}
  \sum_{u\in\mathcal N_r(v)}
  \mathbf m_{u\to v,r}^{(i)},\\
\mathbf q_{v,r}^{(i)}
&=\psi_r^{(i)}\!\left(
  \mathbf a_{v,r}^{\mathrm{att},(i)}
  \Vert\mathbf a_{v,r}^{\mathrm{mean},(i)}\right),
\end{aligned}
\label{eq:relation_aggregation_revised}
\end{equation}
where the aggregates and
$\mathbf q_{v,r}^{(i)}$ belong to $\mathbb R^d$, and
$\psi_r^{(i)}:\mathbb R^{2d}\to\mathbb R^d$ is an \ac{mlp}.
Attention weights sum to one within each destination, relation,
and head before training-time dropout.
Let $\mathcal I_t$ be the relations entering nodes of type $t$.
A type-specific affine map
$U_t^{(i)}:\mathbb R^d\to\mathbb R^d$ processes the node's own state. The type-specific \ac{mlp}
$\eta_t^{(i)}:\mathbb R^{(1+|\mathcal I_t|)d}\to\mathbb R^d$
then combines that state with the relation outputs:
\begin{equation}
\widehat{\mathbf z}_v^{(i)}
=\eta_t^{(i)}\!\left(
U_t^{(i)}(\mathbf z_v^{(i)})\Vert
\mathop{\Vert}_{r\in\mathcal I_t}
\mathbf q_{v,r}^{(i)}\right).
\label{eq:type_specific_relation_fusion}
\end{equation}
An affine map $G_t^{(i)}:\mathbb R^{2d}\to\mathbb R^d$
forms the element-wise gate
$\mathbf g_v^{(i)}
=\sigma\!\left(G_t^{(i)}
(\mathbf z_v^{(i)}\Vert\widehat{\mathbf z}_v^{(i)})\right)
\in(0,1)^d$, where $\sigma$ is the logistic function.
The layer output is
\begin{equation}
\mathbf z_v^{(i+1)}
=\operatorname{LN}_t^{(i)}\!\left(
\mathbf z_v^{(i)}
+\operatorname{Drop}\!\left(
\mathbf g_v^{(i)}
\odot\widehat{\mathbf z}_v^{(i)}\right)\right),
\label{eq:gated_heterogeneous_update_revised}
\end{equation}
where
$\operatorname{LN}_t^{(i)}$ is type-specific layer
normalization. Dropout is disabled during inference. Please refer to Appendix~\ref{app:relation_layer} for the affine parameterizations and training-time dropout. 
At inference, the graph layers are equivariant to consistent relabeling within each node type, with node and edge features permuted accordingly~\cite{Eisen2020REGNN}. 
This claim excludes
the index-dependent direction rule for nearly coincident antennas used by the spacing repair in Section~\ref{subsec:slow_gnn_revised}.

\vspace{-0.5cm}
\subsection{Structured Policy Decoders}
\label{subsec:structured_decoders}

After the relation-aware layers, the slow \ac{ma} states determine the array coordinates. 
The fast \ac{ma} and \ac{ue} states jointly determine each precoder coefficient. Information from target nodes reaches these decoders through the directed graph relations.

\subsubsection{Slow-Timescale Geometry Decoder}
\label{subsec:slow_gnn_revised}

After the $L_{\mathrm s}$ graph layers, a shared position
\ac{mlp} $\phi_{\mathrm R}:\mathbb R^d\to\mathbb R^2$
maps the final \ac{ma} state
$\mathbf z_{\mathrm s,n}^{(L_{\mathrm s})}\in\mathbb R^d$
to a correction relative to
$\mathbf r_n^{\mathrm{ref}}\in\mathbb R^2$.
Let $\mathbf r_{\min},\mathbf r_{\max}\in\mathbb R^2$
be the coordinate bounds of $\mathcal C$. Define
$\mathbf u_n^{\mathrm{ref}}
=\operatorname{clip}_{[\epsilon_r,1-\epsilon_r]}
((\mathbf r_n^{\mathrm{ref}}-\mathbf r_{\min})/
(\mathbf r_{\max}-\mathbf r_{\min}))$ and
$\boldsymbol\ell_n^{\mathrm{ref}}
=\log(\mathbf u_n^{\mathrm{ref}}/
(\mathbf 1_2-\mathbf u_n^{\mathrm{ref}}))$,
where $\epsilon_r\in(0,1/2)$ and the operations are
element-wise. Column $n$ of $\mathbf R^{\mathrm{raw}}$ is
\begin{equation}
\begin{aligned}
\mathbf r_n^{\mathrm{raw}}
&=\mathbf r_{\min}
+(\mathbf r_{\max}-\mathbf r_{\min})
\odot\sigma\!\left(
\boldsymbol\ell_n^{\mathrm{ref}}
+\gamma_{\mathrm R}\phi_{\mathrm R}
(\mathbf z_{\mathrm s,n}^{(L_{\mathrm s})})\right),
\end{aligned}
\label{eq:reference_relative_position_decoder}
\end{equation}
where $\gamma_{\mathrm R}>0$ controls the correction magnitude and $\sigma$ is the element-wise logistic function.
This mapping keeps the raw coordinates within $\mathcal C$.

As will be detailed in Section~\ref{subsec:constrained_training_revised}, training includes a dedicated penalty that encourages separation between \acp{ma}.
Since this penalty does not guarantee feasible spacing for every decoded layout, we additionally apply a finite geometric correction in each forward pass during training and deployment.
The correction acts on antenna pairs with distances below
$D_{\mathrm{proj}}\geq D_{\min}$ before the fast policy and communication-sensing model are evaluated.
Define $\mathcal P=\{(n,\ell):1\leq n<\ell\leq N\}$ and initialize $\mathbf r_n^{(0)}=\mathbf r_n^{\mathrm{raw}}$.
At iteration $j$, let
$d_{n\ell}^{(j)}
=\|\mathbf r_n^{(j)}-\mathbf r_\ell^{(j)}\|_2$.
For $d_{n\ell}^{(j)}>\epsilon_{\mathrm{proj}}$, use
$\widehat{\boldsymbol\delta}_{n\ell}^{(j)}
=(\mathbf r_n^{(j)}-\mathbf r_\ell^{(j)})/d_{n\ell}^{(j)}$.
For nearly coincident positions, use the fixed unit vector
$\operatorname{col}(\cos\theta_{n\ell},\sin\theta_{n\ell})$,
where $\theta_{n\ell}
=2\pi\iota_{n\ell}/(|\mathcal P|+1)$ and
$\iota_{n\ell}$ is the pair index in a fixed ordering.
The proposed pair corrections and simultaneous coordinate update are given by:
\begin{align}
\mathbf p_{n\ell}^{(j)}
&=\tfrac12[D_{\mathrm{proj}}-d_{n\ell}^{(j)}]_+
\widehat{\boldsymbol\delta}_{n\ell}^{(j)},\label{eq:parallel_spacing_projection}\\ \notag
\mathbf r_n^{(j+1)}
&=\Pi_{\mathcal C}\!\Bigg(
\mathbf r_n^{(j)}
+
\tfrac{\eta_{\mathrm p}}{\max\{c_n^{(j)},1\}}
\Bigg[
\sum_{\ell>n}\mathbf p_{n\ell}^{(j)}
-\sum_{\ell<n}\mathbf p_{\ell n}^{(j)}
\Bigg]\Bigg),
\end{align}
where $[x]_+=\max\{x,0\}$,
$\Pi_{\mathcal C}$ clips coordinates to $\mathcal C$,
$\eta_{\mathrm p}\in(0,2]$ is the update scale, and
$c_n^{(j)}$ counts incident pairs whose distance shortfall
exceeds $\epsilon_{\mathrm{proj}}>0$.
Each pair proposes equal and opposite shifts of half the
shortfall, and each antenna averages the incident corrections.
For an isolated violating pair with $d_{n\ell}^{(j)}>\epsilon_{\mathrm{proj}}$
and no boundary clipping, $\eta_{\mathrm p}=1$ reaches the spacing
target in one update. With multiple pairs, simultaneous corrections and clipping can leave residual violations.
The procedure stops when all distances are at least
$D_{\mathrm{proj}}-\epsilon_{\mathrm{proj}}$
or after $I_{\mathrm p}\in\mathbb N$ iterations.
The returned geometry is
$\mathcal P_{\mathrm{MA}}(\mathbf R^{\mathrm{raw}})$.
The fast policy and sensing and rate losses use the repaired geometry, so the forward performance effects of these distance-based corrections are included in joint training.
During training, a \ac{ste} passes gradients to the raw coordinates
\cite{Bengio2013STE}:
\begin{equation}
\mathbf R=\mathbf R^{\mathrm{raw}}
+\operatorname{sg}\!\left[
\mathcal P_{\mathrm{MA}}(\mathbf R^{\mathrm{raw}})
-\mathbf R^{\mathrm{raw}}\right],
\label{eq:straight_through_position_projection_revised}
\end{equation}
where $\operatorname{sg}$ preserves its argument in the forward pass and has zero derivative. 
The resulting identity backward derivative is a surrogate for the derivative of the finite repair.

\begin{remark}[Spacing feasibility]
\label{rem:spacing_feasibility}
Although the finite repair does not provide a universal spacing guarantee, we verified from the unsmoothed coordinates that every repaired geometry checked in Section~\ref{sec:simulation_results}
had a minimum spacing of at least $D_{\min}$.
\end{remark}

\subsubsection{Fast-Timescale Precoder Decoder}
\label{subsec:fast_gnn_revised}

The fast decoder adds learned complex corrections to an
\ac{rzf} reference. For
$\mathbf X\in\mathbb C^{N\times K}$, define
\begin{equation}
\mathcal N_P(\mathbf X)
\triangleq
\frac{\sqrt{P_{\max}}\mathbf X}
{\sqrt{\|\mathbf X\|_{\mathrm F}^2+\epsilon_{\mathrm p}}},
\label{eq:power_normalization_operator_revised}
\end{equation}
where $\epsilon_{\mathrm p}>0$ is a numerical constant.
The resulting transmit power does not exceed $P_{\max}$.
Writing $\mathbf H_b=\mathbf H_b(\mathbf R)$, the reference \ac{rzf} precoder is given by:
\begin{equation}
\begin{aligned}
\mathbf W_b^{\mathrm{RZF}}
&=\mathcal N_P\!\left(
\mathbf H_b
[\mathbf H_b^{\mathsf H}\mathbf H_b+\nu_b\mathbf I_K]^{-1}
\right),\\
\nu_b
&=\gamma_{\mathrm{RZF}}K\overline\sigma_c^2/P_{\max},
\end{aligned}
\label{eq:rzf_reference_precoder_revised}
\end{equation}
where
$\overline\sigma_c^2=K^{-1}\sum_k\sigma_{c,k}^2$
and $\gamma_{\mathrm{RZF}}>0$ controls the regularization.
For each \ac{ma}-\ac{ue} pair $(n,k)$, the shared decoder
$\phi_{\mathrm W}:\mathbb R^{2d+6}\to\mathbb R^2$ takes
the final \ac{ma} and \ac{ue} states, the channel feature
$\mathbf e_{nk,b}^{\mathrm{ch}}\in\mathbb R^3$, and a
three-dimensional feature of the corresponding \ac{rzf}
coefficient. 
The two outputs form the real and imaginary parts
of $[\Delta\mathbf W_b]_{n,k}$. The block precoder is
\begin{equation}
\begin{aligned}
\mathbf W_b
&=\mathcal N_P\!\left(
\mathbf W_b^{\mathrm{RZF}}
+\gamma_{\mathrm W}c_W\Delta\mathbf W_b
\right),\\
c_W&=\sqrt{P_{\max}/(NK)},
\end{aligned}
\label{eq:rzf_residual_precoder_revised}
\end{equation}
where $\gamma_{\mathrm W}>0$ controls the correction magnitude.
The entry-wise corrections can change the \ac{rzf} beam
directions, while final normalization enforces the power
constraint. Please refer to
Appendix~\ref{app:rzf_residual_decoder} for the coefficient
feature and pairwise correction function.

\vspace{-0.5cm}
\subsection{Physics-Informed Constrained Training}
\label{subsec:constrained_training_revised}

At each training step, the selected geometry is evaluated over
$M_{\mathrm{tr}}$ independently sampled block states
$\{\boldsymbol\Xi^{(m)}\}_{m=1}^{M_{\mathrm{tr}}}$.
For each state, the fast policy produces $\mathbf W^{(m)}$,
and the differentiable physics model estimates the angular
\ac{bcrb} $\widehat{\mathcal B}^{(m)}>0$. The sensing loss is
\begin{equation}
\mathcal L_{\mathrm{sen}}
=c_{\mathrm B}\log\!\left(
\max\!\left\{
\frac{1}{M_{\mathrm{tr}}}
\sum_{m=1}^{M_{\mathrm{tr}}}
\widehat{\mathcal B}^{(m)},
\epsilon_{\mathrm B}
\right\}\right),
\label{eq:log_bcrb_loss_revised}
\end{equation}
where $c_{\mathrm B},\epsilon_{\mathrm B}>0$. Above the
numerical floor, the logarithm preserves the ordering of the
sample-average sensing objective while rescaling its gradient.
Each block contributes a gradient through its precoder and the
common geometry.

The nonnegative training margins
$\varepsilon_{\mathrm F},\varepsilon_{\mathrm S}$ and
$\delta_d$ tighten the rate and spacing requirements; evaluation
uses the original thresholds. 
We define $D_{\mathrm{tar}}=D_{\min}+\delta_d$.
The signed
residuals are defined as 
$
g_{k,m}^{\mathrm F}
=
R_k^{\mathrm F,\min}+\varepsilon_{\mathrm F}
  -R_k^{\mathrm{inst},(m)}
$, 
$
g_k^{\mathrm S}
=R_k^{\mathrm S,\min}+\varepsilon_{\mathrm S}
  -\frac{1}{M_{\mathrm{tr}}}
   \sum_{m=1}^{M_{\mathrm{tr}}}
   R_k^{\mathrm{inst},(m)}
$, and
$ 
g_{n\ell}^{d}
=\frac{D_{\mathrm{tar}}
 -\|\mathbf r_n^{\mathrm{raw}}
     -\mathbf r_\ell^{\mathrm{raw}}\|_2}{\lambda}
$. 
Here, $R_k^{\mathrm{inst},(m)}$ is evaluated using the
repaired geometry and sample-$m$ precoder. Nonpositive
residuals satisfy the corresponding tightened requirements.
The rate residuals have units of bit/s/Hz, whereas
$g_{n\ell}^{d}$ is dimensionless.
For a residual $g\in\mathbb R$, multiplier $\upsilon\geq0$, and coefficient $\varrho>0$, we use the one-sided \ac{phr}
term \cite{Rockafellar1974AugmentedLagrangian}
\begin{equation}
\ell_{\varrho}(g,\upsilon)
=\frac{[\upsilon+\varrho g]_+^2-\upsilon^2}{2\varrho},
\qquad [x]_+=\max\{x,0\}.
\label{eq:one_sided_augmented_term_revised}
\end{equation}
Let $\mathcal T_k^{\mathrm F}$ contain the indices of the
$\lceil\tau_{\mathrm F}M_{\mathrm{tr}}\rceil$ largest
$g_{k,m}^{\mathrm F}$, and let
$\mathcal T^d\subseteq\mathcal P$ contain the
$\lceil\tau_d|\mathcal P|\rceil$ largest $g_{n\ell}^{d}$,
where $\tau_{\mathrm F},\tau_d\in(0,1]$. With user-specific
multipliers $\upsilon_k^{\mathrm F},\upsilon_k^{\mathrm S}$
and pair-specific multipliers $\upsilon_{n\ell}^{d}$, all
nonnegative, the constraint losses are
\begin{align}
\mathcal L_{\mathrm F}
&=\frac{1}{K}\sum_{k=1}^{K}
  \frac{1}{|\mathcal T_k^{\mathrm F}|}
  \sum_{m\in\mathcal T_k^{\mathrm F}}
  \ell_{\varrho_{\mathrm F}}
  (g_{k,m}^{\mathrm F},\upsilon_k^{\mathrm F}),\label{eq:constraint_augmented_losses_revised}\\ \notag
\mathcal L_{\mathrm S}
&=\frac{1}{K}\sum_{k=1}^{K}
  \ell_{\varrho_{\mathrm S}}
  (g_k^{\mathrm S},\upsilon_k^{\mathrm S}),\ 
\mathcal L_d
=\frac{1}{|\mathcal T^d|}
  \sum_{(n,\ell)\in\mathcal T^d}
  \ell_{\varrho_d}
  (g_{n\ell}^{d},\upsilon_{n\ell}^{d}).
\end{align}
The positive coefficients
$\varrho_{\mathrm F},\varrho_{\mathrm S},\varrho_d$
control the augmented terms. Tail selection concentrates training on poorly served blocks and closely spaced raw coordinates. 
%
The joint primal loss
$\mathcal L=\mathcal L_{\mathrm{sen}}
+\mathcal L_{\mathrm F}+\mathcal L_{\mathrm S}+\mathcal L_d$
is minimized over
$(\boldsymbol\Theta_{\mathrm s},\boldsymbol\Theta_{\mathrm f})$
using \ac{adamw} with zero weight decay
\cite{Loshchilov2019Decoupled}. After each primal update, define
$\overline g_k^{\mathrm F}
=|\mathcal T_k^{\mathrm F}|^{-1}
\sum_{m\in\mathcal T_k^{\mathrm F}}g_{k,m}^{\mathrm F}$
and $\overline g_k^{\mathrm S}=g_k^{\mathrm S}$.
The multipliers undergo projected signed-residual ascent:
\begin{equation}
\begin{aligned}
\upsilon_k^{\mathrm X}
&\leftarrow
\Pi_{[0,\upsilon_{\max}^{\mathrm X}]}
\bigl(\upsilon_k^{\mathrm X}
+\eta_{\mathrm X}\overline g_k^{\mathrm X}\bigr),\\
\upsilon_{n\ell}^{d}
&\leftarrow
\Pi_{[0,\upsilon_{\max}^{d}]}
\bigl(\upsilon_{n\ell}^{d}
+\eta_d g_{n\ell}^{d}\bigr),
\end{aligned}
\label{eq:projected_dual_updates_revised}
\end{equation}
for $\mathrm X\in\{\mathrm F,\mathrm S\}$, every
$k\in\mathcal K$, and every $(n,\ell)\in\mathcal P$.
Here, $\eta_{\mathrm X},\eta_d>0$ are step sizes,
$\upsilon_{\max}^{\mathrm X},\upsilon_{\max}^{d}>0$
are multiplier caps, and $\Pi_{[a,b]}$ clips its argument
to $[a,b]$. 
All pair multipliers are updated, including pairs outside the current spacing tail. Signed residuals allow multipliers to decrease when their constraints are met with margin. 
The raw-spacing loss encourages separation before repair and the structured decoders enforce the movement region and transmit-power constraints directly.

\vspace{-0.2cm}
\section{Simulation Results}
\label{sec:simulation_results}

In this section, we evaluate the proposed \ac{ttmbgnn} in terms of angular
sensing-error bounds, communication-\ac{qos} satisfaction,
antenna-spacing feasibility, and online computational cost.
All methods use the same test locations, underlying channel draws, reflection coefficients, and target-angle samples, with channels computed for each method's geometry.
The reported root-\ac{bcrb} is
$(180/\pi)\sqrt{N_{\mathrm{test}}^{-1}
\sum_{i=1}^{N_{\mathrm{test}}}\mathcal B_i}$ degrees,
where $N_{\mathrm{test}}$ is the number of test blocks and
$\mathcal B_i$ is the target-averaged angular \ac{bcrb} in radians squared.

\vspace{-0.2cm}
\subsection{Simulation Setup and Implementation}
\label{subsec:simulation_setup}

We consider a monostatic \ac{isac} \ac{bs} at the origin, equipped
with $N=30$ \acp{ma} within the square movement region
$\mathcal C=[-7.5\lambda,7.5\lambda]^2$ on the local $YZ$-plane.
In the default configuration, the \ac{bs} serves $K=3$
single-antenna \acp{ue} and senses $Q=2$ point targets.
We use a $5\times6$ \ac{ura} within $\mathcal C$ as the reference
geometry $\mathbf R^{\mathrm{ref}}$.
In each block, we sample the \ac{ue} and nominal target locations
independently and uniformly with respect to volume within
$\mathcal G_k^{\mathrm{ue}}$ and $\mathcal G_q^{\mathrm{tgt}}$,
respectively.
Channels and target-prior centers vary across blocks, while we
hold the geometry $\mathbf R$ fixed within each slow interval.
Table~\ref{tab:simulation_parameters} summarizes the default principal parameters.
With unity antenna-element and \ac{ue} receive gains and no
additional propagation losses, we compute the communication power
gain as $\beta_{k,b}=(\lambda/(4\pi d_{k,b}^{\mathrm{ue}}))^2$
and the target reflection power as
$|\alpha_{q,b}|^2=\lambda^2\sigma_q/
[(4\pi)^3(d_{q,b}^{\mathrm{tgt}})^4]$.
Here, $d_{k,b}^{\mathrm{ue}}=\|\mathbf p_{k,b}^{\mathrm{ue}}\|_2$
and $d_{q,b}^{\mathrm{tgt}}=\|\mathbf p_{q,b}^{\mathrm{tgt}}\|_2$
are distances from the \ac{bs} reference point, and we set
$\sigma_q=1~\mathrm{m}^2$ for every target.
We generate the complex reflection coefficient as
$\alpha_{q,b}=|\alpha_{q,b}|\exp[\mathrm j
(\phi_{q,b}-4\pi d_{q,b}^{\mathrm{tgt}}/\lambda)]$,
where $\phi_{q,b}\sim\mathcal U[0,2\pi)$ is sampled independently
across targets and blocks.
We regenerate these coefficients across blocks and hold them fixed
within each block, including across all angular-prior samples.
We account for array gains through the steering vectors and
precoders.
In the physics engine, we use $\beta_{k,b}$ to generate channels
and $\alpha_{q,b}$ to evaluate the sensing \ac{fim}.
For the \ac{bcrb}, we treat the real and imaginary parts of
$\alpha_{q,b}$ as unknown nuisance parameters.

For each \ac{qos} pair, we jointly train the slow and fast policies
using uniform per-\ac{ue} requirements.
We estimate the slow statistics from $M_{\mathrm s}=1024$
independent samples and hold these statistics fixed during training.
At each training step, we draw $M_{\mathrm{tr}}=24$ independent
block states and use $S_{\mathrm a}=32$ angular-prior samples
per block.
We arrange the angular samples in $16$ symmetric pairs by
accompanying each sampled offset with its negative about the prior
center.
We select checkpoints using a fixed validation set independent
of the training and final test samples.
For final evaluation, we use $200$ independent trials, each
representing one slow interval of $B=64$ coherence blocks, giving
$N_{\mathrm{test}}=200B=12\,800$ unseen blocks per deterministic
geometry or policy.
Under unchanged statistical inputs, we reuse the geometry
$\mathbf R$ across trials and adapt the precoder $\mathbf W_b$
to each block.
We use $S_{\mathrm a}=96$ angular-prior samples per test block,
generated by the same paired procedure.
We compute $95\%$ confidence intervals from trial-level averages
and apply first-order propagation through the square root for
root-\ac{bcrb}.

\begin{table*}[!t]
\caption{Simulation and Implementation Parameters.
\ac{mlp} widths are listed from input to output.}
\label{tab:simulation_parameters}
\centering
\begingroup
\scriptsize
\renewcommand{\arraystretch}{1.02}
\setlength{\tabcolsep}{2.5pt}
\setlength{\aboverulesep}{0.25ex}
\setlength{\belowrulesep}{0.25ex}
\begin{minipage}[t]{0.493\textwidth}
\vspace{0pt}
\begin{tabularx}{\linewidth}{@{}>{\raggedright\arraybackslash}p{0.235\linewidth}
 >{\raggedright\arraybackslash}X@{}}
\toprule
\multicolumn{2}{c}{\textbf{System and Channel}}\\
\midrule
Dimensions
& \ac{ma}/\ac{ue}/target counts $(N,K,Q)=(30,3,2)$\\
Frequency
& $f_c=28$~GHz,
$\lambda=10.7$~mm, $d_{\mathrm F}=9.6$~m (far-field boundary)\\
Bandwidth / power
& $10$~MHz,
$T=32$ symbols/block, $P_{\max}=30$~dBm\\
Noise
& $\sigma_{\mathrm c}^2=\sigma_{\mathrm s}^2=2.007\times10^{-13}$~W;
$290$~K; communication/sensing noise figures $7$~dB each\\
Array / reference
& $\mathcal C=[-7.5\lambda,7.5\lambda]^2$,
minimum spacing $D_{\min}=0.5\lambda$;
reference: $5\times6$ \ac{ura} over
$[-7.2\lambda,7.2\lambda]^2$\\
Default \acp{ue}
& Centers $\overline{\mathbf p}_k^{\mathrm{ue}}$: $(100,65,20)$, $(100,-65,25)$,
$(125,25,30)$~m; radii $r_k^{\mathrm{ue}}$: $(10,10,12)$~m\\
Added \acp{ue} ($k=4,5$)
& Centers $(111.58,-42.83,22.64)$,
$(119.41,14.80,21.37)$~m; radii $(10,10)$~m\\
Default targets
& Centers $\overline{\mathbf p}_q^{\mathrm{tgt}}$: $(100,0,75)$, $(115,-25,65)$~m;
radii $r_q^{\mathrm{tgt}}$: $(10,10)$~m\\
Added target ($q=3$)
& Center $(107.51,-6.44,67.59)$~m; radius $10$~m\\
Rician / diffuse
& $10\log_{10}\kappa_{k,b}\sim\mathcal U[3,10]$,
$L_{\mathrm N}=4$ clusters, $g_{k,b,\ell}\sim\mathcal{CN}(0,1)$;
spreads $(\sigma_{\mathrm N}^{\mathrm{el}},\sigma_{\mathrm N}^{\mathrm{az}})=(8^\circ,12^\circ)$\\
Target prior / phase
& Raised-cosine full widths $\Delta_q^{\mathrm{el}}=\Delta_q^{\mathrm{az}}=0.20$~rad,
cross section $\sigma_q=1$~m$^2$, $\phi_{q,b}\sim\mathcal U[0,2\pi)$\\
Gains / losses
& All element/\ac{ue} gains $0$~dBi;
system/noise losses $0$~dB\\
\midrule
\multicolumn{2}{c}{\textbf{Network and Numerical Settings}}\\
\midrule
Graph layers
& Node width $d=256$; slow/fast depths $(L_{\mathrm s},L_{\mathrm f})=(3,4)$; $A=4$ attention heads\\
Decoder \acp{mlp}
& Position $\phi_{\mathrm R}$: $256\!\to\!256\!\to\!128\!\to\!2$;
precoder $\phi_{\mathrm W}$: $518\!\to\!256\!\to\!256\!\to\!128\!\to\!2$\\
Output / loss
& Gains $\gamma_{\mathrm R}=\gamma_{\mathrm W}=\gamma_{\mathrm{RZF}}=1$;
clipping $\epsilon_r=10^{-4}$, power/log floors
$\epsilon_{\mathrm p}=\epsilon_{\mathrm B}=10^{-12}$, sensing weight $c_{\mathrm B}=1$\\
\bottomrule
\end{tabularx}
\end{minipage}\hfill
\begin{minipage}[t]{0.493\textwidth}
\vspace{0pt}
\begin{tabularx}{\linewidth}{@{}>{\raggedright\arraybackslash}p{0.235\linewidth}
 >{\raggedright\arraybackslash}X@{}}
\toprule
\multicolumn{2}{c}{\textbf{Training and Evaluation}}\\
\midrule
\multicolumn{2}{@{}l@{}}{Node/relation \acp{mlp}: two affine layers, hidden width $256$, SiLU.}\\
Optimizer / budget
& \ac{adamw}, learning rate $7.5\times10^{-5}$,
$200$ epochs, $100$ steps/epoch\\
Input scaling
& Gain offset $c_\beta=10$, Rician scale $c_\kappa=10$;
rate scale $c_R=10$~bit/s/Hz; dropout $p_{\mathrm{drop}}=0.05$\\
Clipping / decay
& Slow/fast gradient-norm caps $0.5/1$; weight decay $0$\\
\ac{qos} pairs
& $(R^{\mathrm F,\min},R^{\mathrm S,\min})
\in\{(1,2),\ldots,(6,7)\}$~bit/s/Hz\\
Margins / tails
& $(\varepsilon_{\mathrm F},\varepsilon_{\mathrm S})=(0.25,0.10)$~bit/s/Hz;
$(\tau_{\mathrm F},\tau_d)=(0.25,0.10)$\\
\ac{phr} / dual steps
& $(\varrho_{\mathrm F},\varrho_{\mathrm S},\varrho_d)=(10,10,20)$;
$(\eta_{\mathrm F},\eta_{\mathrm S},\eta_d)=(0.01,0.01,0.02)$\\
Multiplier values
& Initially $0$;
$(\upsilon_{\max}^{\mathrm F},\upsilon_{\max}^{\mathrm S},\upsilon_{\max}^{d})=(50,50,100)$\\
Spacing repair
& Targets $D_{\mathrm{tar}}=0.52\lambda$,
$D_{\mathrm{proj}}=0.521\lambda$; $I_{\mathrm p}=96$ iterations,
step $\eta_{\mathrm p}=1$, tolerance $\epsilon_{\mathrm{proj}}=10^{-5}\lambda$\\
Training / validation
& $M_{\mathrm s}=1024$ statistics samples; training: $M_{\mathrm{tr}}=24$ blocks, $S_{\mathrm a}=32$ angles/block;
validation: $256$ blocks, $64$ angular samples/block, every epoch\\
Final test
& $200$ slow intervals, $B=64$ blocks/interval,
$96$ angular samples/block\\
Size adaptation
& $10$ epochs, $100$ steps/epoch;
learning rate $2.5\times10^{-5}$\\
\midrule
\multicolumn{2}{c}{\textbf{\ac{saa-ao}} (Adam, learning rate $0.02$)}\\
\midrule
Samples / updates
& $64$ blocks, $48$ angular samples/block;
$30$ outer iterations, each with $5$ geometry and $30$ precoder updates\\
Initialization
& $20$ starts: reference array and $19$ layouts with independent
$\mathcal N(0,0.25^2)$ perturbations of coordinate logits;
\ac{rzf} precoders\\
Primal updates
& Geometry/precoder gradient-norm caps $1/5$\\
Constraint terms
& Rate margins $0$; fast/slow/spacing penalties $(10,10,20)$ and
dual steps $(0.01,0.01,0.02)$; multipliers initially $0$, cap $100$\\
Test refinement
& $100$ precoder updates, gradient-norm cap $5$, penalty coefficient $10$;
dual step $0.05$, initial value $0$, cap $100$\\
\bottomrule
\end{tabularx}
\end{minipage}
\endgroup
\vspace{-0.5cm}
\end{table*}

\vspace{-0.6cm}
\subsection{Benchmarks and Ablation Studies}
\label{subsec:simulation_benchmarks}

We compare two learned-component ablations and three independent
baselines on common test realizations under the same physical and
communication requirements.

\subsubsection{Learned-Component Ablations}
We use the following variants without retraining their learned
components:
\begin{itemize}
\setlength{\itemsep}{1pt}\setlength{\parskip}{0pt}
\item \emph{Fixed \ac{ma}+\ac{fmbgnn}:}
We replace the learned positions with the feasible reference
geometry to assess the contribution of geometry adaptation.

\item \emph{\ac{smbgnn}+\ac{rzf}:}
In this baseline, we combine the learned geometry with blockwise \ac{rzf} to assess
the contribution of learned precoding.
\end{itemize}

\subsubsection{Independent Baselines}
\begin{itemize}
\setlength{\itemsep}{1pt}\setlength{\parskip}{0pt}
\item \emph{Fixed \ac{ma}+\ac{rzf}:}
We combine the reference geometry with blockwise \ac{rzf} as a
nonlearning benchmark.

\item \emph{\ac{saa-ao}:}
We approximate the long-term objective using finite sample averages
and alternately update the common geometry and sample-dependent
precoders from multiple
initializations~\cite{Ma2026MAISAC_StatCSI}.
We also perform iterative precoder refinement on test blocks.
The optimizer uses simulated target reflection coefficients to
evaluate the sensing objective throughout optimization and refinement.
In the proposed framework, we use these nuisance values only for
offline loss computation and performance evaluation.
Thus, \ac{saa-ao} serves as an idealized benchmark with additional
sensing information and iterative computation at deployment.

\item \emph{Fixed \ac{ma}+sensing-only eigenbeams:}
For the reference geometry, we use the $K$ leading eigenvectors
of the steering-vector covariance averaged over samples from the
current target-angle priors and normalize the precoder to the
power budget.
We include this sensing-oriented diagnostic without communication
\ac{qos} constraints or direct angular \ac{bcrb} minimization.
\end{itemize}

\vspace{-0.3cm}
\subsection{Training Convergence}
\label{subsec:simulation_convergence}

\begin{figure*}[!t]
    \centering
    \includegraphics[
        width=\textwidth,
        pagebox=cropbox,
        clip
    ]{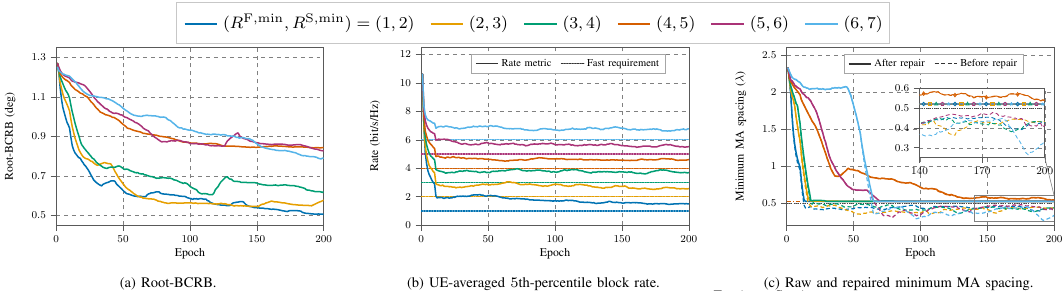}
    \caption{Fixed-validation convergence of the proposed
    \ac{ttmbgnn} under uniform per-\ac{ue} \ac{qos} pairs
    $(R^{\mathrm F,\min},R^{\mathrm S,\min})$.
    In panel~(b), solid curves show the \ac{ue}-averaged
    $5$th-percentile block rates, and dotted horizontal lines denote
    the fast-timescale requirements.
    In panel~(c), solid and dashed curves show repaired and raw
    minimum spacings, respectively. The inset highlights several
    repaired curves approaching $D_{\mathrm{proj}}=0.521\lambda$,
    with staggered markers distinguishing their overlap.
    The physical spacing requirement is $D_{\min}=0.5\lambda$.
    For metric curves, we use a ten-epoch trailing average for
    better visualization.}
    \label{fig:training_convergence}
\vspace{-0.5cm}
\end{figure*}

Fig.~\ref{fig:training_convergence} presents the fixed-validation
evolution under the six \ac{qos} configurations.
The displayed curves use a ten-epoch trailing average for readability,
whereas feasibility checks and final evaluation use unsmoothed metrics.
In Fig.~\ref{fig:training_convergence}a, the root-\ac{bcrb} decreases rapidly
and then stabilizes within a \ac{qos}-dependent range.
Residual fluctuations and the lack of strict ordering across independently
trained models reflect stochastic sampling and nonconvex primal-dual
optimization. More stringent communication requirements generally
yield larger sensing bounds.
Fig.~\ref{fig:training_convergence}b shows that the \ac{ue}-averaged
$5$th-percentile block rate settles above the corresponding fast-timescale requirement as the initially large \ac{rzf}-based rate margin decreases and sensing performance improves.
This aggregate percentile does not establish strict per-block feasibility. 
The maximum per-\ac{ue} slow-rate shortfall was zero at every recorded validation epoch and is omitted.
Fig.~\ref{fig:training_convergence}c shows that the raw minimum spacing can
fall below $D_{\min}$, while the repaired curves remain close to
$D_{\mathrm{proj}}$.
This observed behavior demonstrates the empirical effectiveness
of the repair without establishing a universal spacing guarantee.
Physical feasibility is assessed from the unsmoothed repaired
coordinates against $D_{\min}$.
The raw-spacing augmented-Lagrangian term separately discourages
overlapping decoder outputs.
\vspace{-0.4cm}

\subsection{Sensing-Communication Tradeoff and \ac{qos} Reliability}
\label{subsec:simulation_tradeoff}

\begin{figure}[!t]
    \centering
    \includegraphics[
        width=\columnwidth,
        pagebox=cropbox,
        clip
    ]{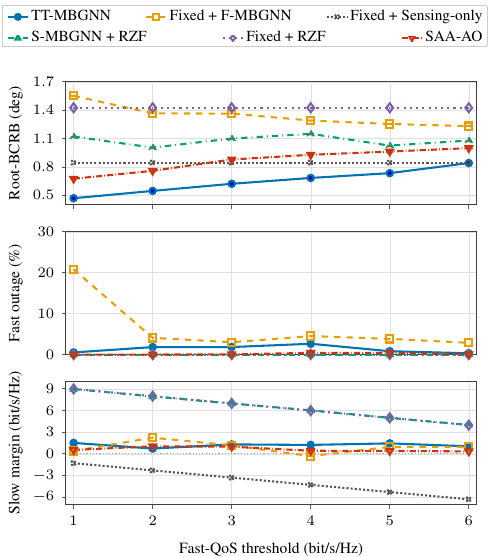}
    \caption{Root-\ac{bcrb} (top), any-\ac{ue} fast-\ac{qos}
    outage (middle), and minimum per-\ac{ue} slow-rate margin
    (bottom) versus the uniform fast-\ac{qos} threshold.
    The slow threshold is
    $R^{\mathrm S,\min}=R^{\mathrm F,\min}+1~\mathrm{bit/s/Hz}$.
    The sensing-only design is omitted from the middle panel
    because its fast outage exceeds $98\%$ at every threshold.
    The dotted horizontal line denotes zero slow-rate margin.}
    \label{fig:qos_sensing_tradeoff}
\end{figure}

Fig.~\ref{fig:qos_sensing_tradeoff} reports results on unseen test
blocks. The fast-rate threshold varies from $1$ to
$6$~bit/s/Hz, and the slow-rate threshold is $1$~bit/s/Hz higher.
The middle panel reports the fraction of blocks with a fast-rate
violation by any \ac{ue} defined as 
$
\widehat p_{\mathrm F}
=\frac{1}{N_{\mathrm{test}}}\sum_i
\mathbb I\!\{\exists k:
R_{k,i}^{\mathrm{inst}}<R_k^{\mathrm F,\min}\}
$. 
The bottom panel shows
$s_{\mathrm S}=\min_k(\widehat{\overline R}_k-R_k^{\mathrm S,\min})$,
where $\widehat{\overline R}_k$ is the test-mean rate of \ac{ue} $k$.
These empirical checks are applied to each method and threshold pair.
A sensing-bound value alone does not establish \ac{qos} feasibility and should be interpreted alongside the communication results.
The proposed \ac{ttmbgnn} achieves a lower root-\ac{bcrb} than
both learned-component ablations throughout the sweep, supporting
joint geometry and precoder adaptation.
We observe positive slow-rate margins and low, nonzero fast outage
for the proposed method.
The remaining outages indicate that some test blocks violate the
fast-rate requirements.
The analytical \ac{rzf} designs show no observed fast outage and
larger slow-rate margins, together with higher sensing bounds.
The sensing-only eigenbeams exhibit $98$--$100\%$ fast outage and
negative slow-rate margins.
We omit their outage curve from the middle panel to improve the
visibility of the other methods.

\vspace{-0.3cm}
\subsection{Generalization Across Network Sizes}
\label{subsec:simulation_size_generalization}
\begin{figure}[!t]
    \centering
    \includegraphics[
        width=\columnwidth,
        pagebox=cropbox,
        clip
    ]{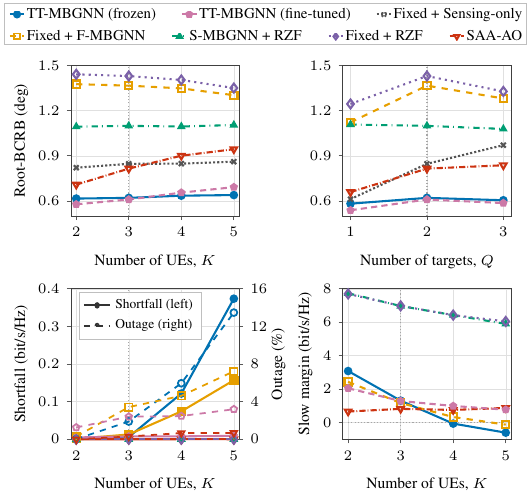}
    \caption{Network-size transfer from $(K,Q)=(3,2)$ with
    frozen parameters or ten fine-tuning epochs.
    Top: root-\ac{bcrb} versus $K$ ($Q=2$) and $Q$ ($K=3$).
    Bottom ($Q=2$): mean worst-\ac{ue} fast-rate shortfall
    (solid, left axis) and any-\ac{ue} fast-\ac{qos} outage
    (dashed, right axis) in the left panel, and minimum slow-rate
    margin in the right panel.
    Dotted lines mark the training size and zero.
    \emph{Fixed} denotes the reference array.
    Sensing-only beams appear only in the top row.}
    \label{fig:size_transfer}
\end{figure}

\vspace{-0.1cm}
Fig.~\ref{fig:size_transfer} evaluates policies trained with $(K,Q)=(3,2)$ and uniform fast and slow \ac{qos} requirements of $3$ and $4$~bit/s/Hz, respectively. 
Additional-node centers and region radii are specified separately in Table~\ref{tab:simulation_parameters}. The top panels show the root-\ac{bcrb} as $K$ or $Q$ varies. 
The bottom-left panel reports the test-block mean worst-\ac{ue} fast-rate shortfall,
$\widehat{\mathbb E}_{\mathrm{test}}
[\max_k[R_k^{\mathrm F,\min}-R_{k,b}^{\mathrm{inst}}]_+]$, on the left axis and the any-\ac{ue} fast-\ac{qos} outage percentage on the right axis. The latter is the percentage of test blocks in which at least one \ac{ue} misses the fast-rate requirement.
The bottom-right panel reports the minimum slow-rate margin,
$\min_k(\widehat{\overline R}_k-R_k^{\mathrm S,\min})$.

The frozen \ac{ttmbgnn} achieves favorable sensing bounds across the tested numbers of \acp{ue} and targets.
Increasing $K$ beyond the training size raises the communication load, resulting in larger fast-rate shortfalls, more frequent outages, and eventually a negative slow-rate margin. 
Ten fine-tuning epochs, corresponding to $5\%$ of the original training epoch count, substantially reduce fast-rate shortfalls and outage at these larger user counts while maintaining favorable sensing bounds. 
The fine-tuned policies also yield positive slow-rate margins across all tested user counts.
The remaining nonzero fast outage indicates residual block-level violations. These results support reuse of the trained policies with limited adaptation to increased communication load.

\vspace{-0.5cm}
\subsection{Online Computational Cost}
\label{subsec:simulation_complexity}

Table~\ref{tab:online_latency} reports batch-one precoder-update
latency at $(N,K,Q)=(30,5,2)$, measured on an NVIDIA GeForce
RTX 5080 GPU (16~GB) with an Intel Core Ultra~9 275HX CPU.
The proposed \ac{ttmbgnn} requires $16.48$~ms per block,
compared with $832.80$~ms for \ac{saa-ao} with $100$ updates,
giving a $50.5$-fold speedup.
These timings exclude channel acquisition and antenna movement.
The speedup depends on network size, hardware, and iteration
budget, while the proposed fast policy requires no online
optimization loop.

\begin{table}[!t]
\caption{Median Batch-One Fast-Update Latency}
\label{tab:online_latency}
\centering
\footnotesize
\renewcommand{\arraystretch}{1.10}
\setlength{\tabcolsep}{5pt}
\scalebox{0.8}{%
\begin{tabular}{@{}lcc@{}}
\toprule
Method & Proposed \ac{ttmbgnn} & \ac{saa-ao} ($100$ updates) \\
\midrule
Median (ms) & 16.48 & 832.80 \\
\bottomrule
\end{tabular}%
}
\end{table}

\vspace{-0.3cm}
\section{Conclusion}
\label{sec:conclusion}

In this paper, we developed a two-timescale \ac{ma}-assisted
\ac{isac} framework with block-level and long-term per-\ac{ue}
rate requirements.
The proposed \ac{ttmbgnn} combines an \ac{smbgnn} for antenna
positioning with an \ac{fmbgnn} for block-adaptive precoding.
Relation-aware layers represent interactions among antennas,
users, and targets.
A bounded position decoder is followed by finite spacing repair,
with a \ac{ste} supplying surrogate gradients during training.
The fast decoder learns complex corrections around an \ac{rzf} reference.
Joint training combines a Monte Carlo angular \ac{bcrb} objective with one-sided augmented-Lagrangian terms for rate requirements
and raw-spacing violations.
Simulations demonstrated favorable sensing-communication tradeoffs
relative to the principal ablations and analytical baselines.
The proposed method achieved competitive sensing bounds against
the idealized \ac{saa-ao} benchmark and approximately $50$-fold faster batch-one precoder computation under the reported settings.
Frozen policies provided competitive sensing bounds at the tested unseen user and target counts, while brief fine-tuning improved communication reliability at larger user counts.
Communication reliability was assessed empirically, with residual fast-rate outages.
Every repaired geometry checked in the reported simulations met the minimum antenna spacing requirement, although the finite repair provides no universal spacing guarantee.
Future work will examine robustness to imperfect \ac{csi} and array calibration errors and validate the framework on practical \ac{ma} platforms.

\appendices

\vspace{-0.4cm}
\section{Input-Feature Construction}
\label{app:feature_construction}

This appendix gives the explicit construction and normalization of the
feature collections in
\eqref{eq:slow_feature_map_compact} and
\eqref{eq:fast_feature_map_compact}. Each node-feature vector below is
a column vector whose transpose forms the corresponding row of its
feature matrix. For example,
$[\mathbf X_{\mathrm s}^{\mathrm{ma}}]_{n,:}
=(\mathbf x_n^{\mathrm s,\mathrm{ma}})^{\mathsf T}$.
The same convention applies to the remaining slow- and fast-timescale
features. The vectors
$\mathbf r_n^{\mathrm{ref}},\mathbf r_n\in\mathbb R^2$
are columns of $\mathbf R^{\mathrm{ref}}$ and $\mathbf R$,
respectively.

\vspace{-0.4cm}
\subsection{Slow-Timescale Features}

The reference position is normalized to the movement region as
\begin{equation}
\mathbf x_n^{\mathrm s,\mathrm{ma}}
=2\tfrac{\mathbf r_n^{\mathrm{ref}}-\mathbf r_{\min}}
{\mathbf r_{\max}-\mathbf r_{\min}}-\mathbf 1_2
\in\mathbb R^2.
\label{eq:slow_ma_feature_appendix}
\end{equation}
For \ac{ue} $k$ and $\nu\in\{\mathrm{el},\mathrm{az}\}$, let
$\vartheta_{k,b}^{\mathrm{LoS},\nu}$ be the corresponding
line-of-sight departure angle and define its circular mean by
\begin{equation}
\begin{aligned}
\overline\vartheta_k^\nu
=
\operatorname{atan2}\!\big(&
\mathbb E[\sin\vartheta_{k,b}^{\mathrm{LoS},\nu}],
\mathbb E[\cos\vartheta_{k,b}^{\mathrm{LoS},\nu}]\big).
\end{aligned}
\label{eq:slow_circular_mean_direction}
\end{equation}
Let $\beta_{k,b}>0$ denote the large-scale channel power gain and
$\kappa_{k,b}\geq0$ the Rician factor in linear units. Their scaled
statistics are
$\widetilde\beta_k=\log_{10}(\mathbb E[\beta_{k,b}])+c_\beta$
and $\widetilde\kappa_k=\mathbb E[\kappa_{k,b}]/c_\kappa$,
where $c_\beta\in\mathbb R$ is a fixed offset and $c_\kappa>0$
is a scale. With rate scale $c_R>0$ in bit/s/Hz, the feature is
\begin{equation}
\begin{aligned}
\mathbf x_k^{\mathrm s,\mathrm{ue}}
=\operatorname{col}\!\big(&
\sin\overline\vartheta_k^{\mathrm{el}},
\cos\overline\vartheta_k^{\mathrm{el}},\\
&\sin\overline\vartheta_k^{\mathrm{az}},
\cos\overline\vartheta_k^{\mathrm{az}},
\widetilde\beta_k,\widetilde\kappa_k,
R_k^{\mathrm{S},\min}/c_R\big)
\in\mathbb R^7.
\end{aligned}
\label{eq:slow_ue_feature}
\end{equation}
The expectations are computed offline from the long-term
distributions using \eqref{eq:general_sample_average}.
For target $q$, define the nominal angle
$\mu_{q,\mathrm S}^{\nu}
=\vartheta^{\nu}(\overline{\mathbf p}_q^{\mathrm{tgt}})$,
where $\vartheta^\nu(\cdot)$ is the Cartesian-to-angle map used
in the system model. The angular widths are $\Delta_q^\nu>0$.
The periodic feature
\begin{equation}
\begin{aligned}
\mathbf x_q^{\mathrm s,\mathrm{tgt}}
=\operatorname{col}\!\big(&
\sin\mu_{q,\mathrm S}^{\mathrm{el}},
\cos\mu_{q,\mathrm S}^{\mathrm{el}},\\
&\sin\mu_{q,\mathrm S}^{\mathrm{az}},
\cos\mu_{q,\mathrm S}^{\mathrm{az}},
\Delta_q^{\mathrm{el}}/\pi,
\Delta_q^{\mathrm{az}}/\pi\big)
\in\mathbb R^6
\end{aligned}
\label{eq:slow_target_periodic_feature}
\end{equation}
avoids discontinuities from angular wrapping
\cite{Mardia2000DirectionalStatistics}.

\vspace{-0.4cm}
\subsection{Fast-Timescale Features}

For the fast \ac{ma} feature, we use the selected geometry:
\begin{equation}
\mathbf x_{n,b}^{\mathrm f,\mathrm{ma}}
=2\tfrac{\mathbf r_n-\mathbf r_{\min}}
{\mathbf r_{\max}-\mathbf r_{\min}}-\mathbf 1_2
\in\mathbb R^2.
\label{eq:fast_ma_feature_corrected}
\end{equation}
This vector is constant within one positioning interval but generally
differs from the reference feature in
\eqref{eq:slow_ma_feature_appendix}. 
For the target's features, we form the following using the prior:
\begin{equation}
\begin{aligned}
\mathbf x_{q,b}^{\mathrm f,\mathrm{tgt}}
=\operatorname{col}\!\big(&
\sin\mu_{q,b}^{\mathrm{el}},\cos\mu_{q,b}^{\mathrm{el}},\\
&\sin\mu_{q,b}^{\mathrm{az}},\cos\mu_{q,b}^{\mathrm{az}},
\Delta_{q,b}^{\mathrm{el}}/\pi,
\Delta_{q,b}^{\mathrm{az}}/\pi\big)
\in\mathbb R^6.
\end{aligned}
\label{eq:fast_target_periodic_feature}
\end{equation}

For channel coefficient $h_{n,k,b}=[\mathbf H_b]_{n,k}\in\mathbb C$,
define $\check h_{n,k,b}=h_{n,k,b}\sqrt{P_{\max}/\sigma_{c,k}^2}$
and the real-valued compression
$\operatorname{slog}(x)=\operatorname{sign}(x)\log(1+|x|)$.
The channel feature is
\begin{equation}
\begin{aligned}
\mathbf e_{nk,b}^{\mathrm{ch}}
=\operatorname{col}\!\big(&
\operatorname{slog}(\Re\{\check h_{n,k,b}\}),\\
&\operatorname{slog}(\Im\{\check h_{n,k,b}\}),
\log(1+|\check h_{n,k,b}|)\big)
\in\mathbb R^3.
\end{aligned}
\label{eq:channel_edge_feature}
\end{equation}
The tensor entry $[\mathbf E_b^{\mathrm{ch}}]_{n,k,:}$ stores this
vector. 
Both directed computational edges for pair $(n,k)$ use the same
feature. Reversing a graph edge does not conjugate the physical
channel coefficient.
With $s_{n,k,b}^{\mathrm{mag}}=\log(1+|\check h_{n,k,b}|)$,
the \ac{ue} feature is
\begin{equation}
\begin{aligned}
\mathbf x_{k,b}^{\mathrm f,\mathrm{ue}}
&=\operatorname{col}\!\big(
\tfrac{1}{N}\textstyle\sum_{n=1}^{N}s_{n,k,b}^{\mathrm{mag}},
\sqrt{\tfrac{1}{N}\textstyle\sum_{n=1}^{N}(s_{n,k,b}^{\mathrm{mag}})^2},\\
&\hspace{12mm}\max_n s_{n,k,b}^{\mathrm{mag}},
R_k^{\mathrm{F},\min}/c_R\big)
\in\mathbb R^4.
\end{aligned}
\label{eq:fast_ue_feature}
\end{equation}
The node-level channel summaries complement the link-level complex
channel information in $\mathbf E_b^{\mathrm{ch}}$. Together,
\eqref{eq:fast_ma_feature_corrected},
\eqref{eq:fast_target_periodic_feature},
\eqref{eq:channel_edge_feature}, and
\eqref{eq:fast_ue_feature} provide the entries of
$\mathcal X_{\mathrm f,b}$ defined in
\eqref{eq:fast_feature_map_compact}.

\vspace{-0.3cm}
\section{Relation-Aware Layer Implementation}
\label{app:relation_layer}

In \eqref{eq:relation_attention_weights}, the score matrix
$\mathbf C_r^{(i)}\in\mathbb R^{A\times d}$ has no bias. 
Each
$d$-dimensional message is divided into $A$ consecutive heads of dimension $d_h=d/A$. Attention weights are normalized over the source nodes of the same relation and destination. 
During training, dropout is applied to these weights before the attention aggregate in \eqref{eq:relation_aggregation_revised} is formed; the mean aggregate is computed from the original messages. Attention dropout
is disabled during inference.
The affine self transformation and gate map used in
\eqref{eq:type_specific_relation_fusion} and
\eqref{eq:gated_heterogeneous_update_revised} are given by 
$
U_t^{(i)}(\mathbf z)
=\mathbf W_t^{\mathrm{self},(i)}\mathbf z
  +\mathbf b_t^{\mathrm{self},(i)},\ 
G_t^{(i)}(\mathbf z\Vert\widehat{\mathbf z})
=\mathbf W_t^{g,(i)}
  [\mathbf z\Vert\widehat{\mathbf z}]
  +\mathbf b_t^{g,(i)}.
$
Here,
$\mathbf W_t^{\mathrm{self},(i)}\in\mathbb R^{d\times d}$,
$\mathbf W_t^{g,(i)}\in\mathbb R^{d\times 2d}$, and both bias
vectors belong to $\mathbb R^d$. The gate is
$\mathbf g_v^{(i)}
=\sigma\!\left(
G_t^{(i)}(\mathbf z_v^{(i)}
\Vert\widehat{\mathbf z}_v^{(i)})
\right)$.
The message function $\phi_r^{(i)}$, relation fusion function
$\psi_r^{(i)}$, and type-specific fusion function $\eta_t^{(i)}$
each use two affine layers with a $d$-dimensional hidden layer,
a SiLU activation, and hidden-layer dropout. Relations entering
a node are concatenated in a fixed order. The \ac{ma} and
\ac{ue} nodes each receive two relation types, whereas target
nodes receive one. Parameters are shared among nodes of the
same type within a policy, but are distinct across layers and between the slow and fast policies.

\vspace{-0.3cm}
\section{Pairwise \ac{rzf}-Residual Precoder Decoder}
\label{app:rzf_residual_decoder}

Normalize the reference coefficient as
$\widetilde w_{n,k,b}^{\mathrm{RZF}}
=[\mathbf W_b^{\mathrm{RZF}}]_{n,k}/c_W\in\mathbb C$.
Using the compression defined in
Appendix~\ref{app:feature_construction}, form
\begin{equation}
\begin{aligned}
\mathbf e_{nk,b}^{\mathrm{RZF}}
=\operatorname{col}\!\big(&
\operatorname{slog}(\Re\{\widetilde w_{n,k,b}^{\mathrm{RZF}}\}),\\
&\operatorname{slog}(\Im\{\widetilde w_{n,k,b}^{\mathrm{RZF}}\}),
\log(1+|\widetilde w_{n,k,b}^{\mathrm{RZF}}|)\big)
\in\mathbb R^3.
\end{aligned}
\label{eq:rzf_coefficient_feature}
\end{equation}
The shared \ac{mlp}
$\phi_{\mathrm W}:\mathbb R^{2d+6}\to\mathbb R^2$ produces
the real correction vector $\boldsymbol\delta_{n,k,b}\in\mathbb R^2$:
\begin{equation}
\begin{aligned}
\boldsymbol\delta_{n,k,b}
&=\phi_{\mathrm W}\!\big(
\mathbf z_{\mathrm f,n,b}^{\mathrm{ma},(L_{\mathrm f})}
\Vert\mathbf z_{\mathrm f,k,b}^{\mathrm{ue},(L_{\mathrm f})}
\Vert\mathbf e_{nk,b}^{\mathrm{ch}}
\Vert\mathbf e_{nk,b}^{\mathrm{RZF}}\big),\\
[\Delta\mathbf W_b]_{n,k}
&=[\boldsymbol\delta_{n,k,b}]_1
+\mathrm j[\boldsymbol\delta_{n,k,b}]_2,
\end{aligned}
\label{eq:precoder_pair_decoder}
\end{equation}
Scaling and normalization then follow
\eqref{eq:rzf_residual_precoder_revised}.

\vspace{-0.3cm}
\bibliographystyle{IEEEtran}
\bibliography{IEEEabrv,Bib}
\end{document}